\documentclass[reprint,twocolumn,showpacs,pre,nofootinbib,amsmath,amssymb,aps,floatfix]{revtex4-2}
\usepackage{graphicx} 
\usepackage{dcolumn}
\usepackage{bm}
\usepackage{amsmath}
\usepackage{epsfig}
\usepackage{color,hyperref}

\usepackage{appendix}

\begin{document}

\title{Occupation time statistics for non-Markovian random walks}

%\author{V. M\'endez$^{1}$, R. Flaquer-Galm\'es$^{1}$, A. Pal$^{2,3}$}
\author{V. M\'endez$^{1}$}
\email{vicenc.mendez@uab.cat}
\author{R. Flaquer-Galm\'es$^{1}$}
\author{A. Pal$^{2,3}$}
\email{arnabpal@imsc.res.in}

\affiliation{$^{1}$Grup de F\'{\i}sica Estad\'{\i}stica, Departament de F\'{\i}sica. Facultat de Ci\`{e}ncies, Universitat Aut\`{o}noma de Barcelona, 08193 Barcelona, Spain.\\
$^{2}$ The Institute of Mathematical Sciences, CIT Campus, Taramani, Chennai 600113, India.\\
$^{3}$ Homi Bhabha National Institute, Training School Complex, Anushakti Nagar, Mumbai 400094}

\begin{abstract}
We study the occupation time statistics for non-Markovian random walkers based on the formalism of the generalized master equation for the Continuous-Time Random Walk. We also explore the case when the random walker additionally undergoes a stochastic resetting dynamics. We derive and solve the backward Feynman-Kac equation to find the characteristic function for the occupation time in an interval and for the half occupation time in the semi-infinite domain. We analyze the behaviour of the PDFs, the moments, the limiting distributions and the ergodic properties for both occupation times when the underlying random walk is normal or anomalous. For the half occupation time, we revisit the famous arcsine law and examine its validity pertaining to various regimes of the rest period of the walker. Our results have been verified with numerical simulations exhibiting an excellent agreement.    
\end{abstract}

\maketitle

\section{Introduction}
\label{Introduction}
The study of animal movement, particularly concerning foraging behavior, provides insights into ecological dynamics and animal movement patterns. In recent years, the monitoring and subsequent analysis of animal movements or trajectories connected to statistical physics and dynamical systems has gained attention as an effective field of study \cite{Fe16}. The most simple approach to describe the behavioral navigation of living beings is to consider the walker's position in time as a Markovian (i.e., a memory-less) stochastic process \cite{CaMe19}. However, animals employ various strategies to explore their environment efficiently, often influenced by memory and other cognitive mechanisms. For instance, it has been recurrently observed for many species the tendency to repeatedly return to previously visited sites as a part of their regular foraging strategies \cite{Vi22}. For this reason, incorporating memory within random walk models is important for improving both predictive and descriptive movement tools. This can be done, for example, by considering the resetting of the walker's position to a given point \cite{EvMaSh20} (spatial memory) or long rests between successive jumps (temporal memory) \cite{Vi22b}. In both cases, the resulting random walk is non-Markovian. 

To extract statistical information from these non-Markovian models, one can analyze stochastic functionals of the walker's trajectory. The prototypical example is the Brownian functional, an integration of the trajectory of a Brownian walker \cite{Ma05,majumdar2020statistics}. They have diverse applications in physics, mathematics, and other fields. The probability density function (PDF) of Brownian functionals satisfies the Feynman-Kac (FK) equation, which is a Schr\"odinger equation in imaginary time. This equation was derived in 1949 by Kac inspired by Feynman’s path integrals. More recently the FK equation has been derived for resetting Brownian walkers \cite{Ho19,singh2022first,dubey2023first,pal2019local,yan2023breakdown} and non-Brownian subdiffusion \cite{TuCaBa09,CaBa10}. However, derivation for a general FK equation for any Markovian or non-Markovian random walk is still missing.

Several functionals are of ecological interest. Take e.g., the occupation time or residence time, which estimates the time spent by a random walker in a given domain which has ecological relevance. If $x=0$ is the starting point and the animal's nest location then the interval $[-a,a]$ around $x=0$ can be regarded as the home range and the occupation time in $[-a,a]$ can be interpreted as the time spent by the animal within such home range. It provides useful information on the animal's behavior when foraging. One can also consider the half occupation time (the occupation time in the domain $x>0$) or the occupation time in a finite interval which can provide information of a restricted foraging space. The distribution function for the half occupation time of a Brownian walker is given by the seminal L\'evy's arcsine law \cite{Le40,majumdar2002local,sabhapandit2006statistical,burenev2024occupation,kay2023extreme} and holds a special place in stochastic process, probability theory and physics. Statistics of occupation times are of course not limited to Brownian walker, and it is a topic of wide investigation for example in the context of renewal processes theory \cite{GoLu01}, active particles \cite{Br20,singh2019generalised}, fractional Brownian motion \cite{sadhu2018generalized}, experiments on blinking quantum dots \cite{MaBa05}, colloids modulated with optical traps \cite{dey2022experimental} and light resonators in a cavity \cite{ramesh2024arcsine} and weak ergodicity breaking of dynamics generated using deterministic maps \cite{BeBa06}. On the other hand, the occupation time in an interval for a subdiffusive walker has been studied in Ref. \cite{CaBa10}. More recently, the statistical and ergodic properties of a Brownian walker in the presence of power-law resetting have been studied \cite{BaFlMe24}.  Despite these studies, a unified framework analyzing the occupation time spanned in different geometries for generic random walkers is missing.

The aim of this work is to fill in this gap by developing a comprehensive framework to study the occupation time in a confinement and semi-confinement for generic random walkers namely the Markovian- and non-Markovian- walkers. To this end, we derive the Feynman-Kac equation for the occupation time of generic random walkers with arbitrary rest periods. We solve for the characteristic equation, compute the moments, obtain the PDF, analyze the ergodic properties, and extend our study considering the stochastic resetting of the walker's position. The article is organized as follows. In Section \ref{CTRW-formalism} we review the derivation of the Generalized Master equation for the Continuous Time Random Walk (CTRW). In Section \ref{FK-formalism}, we derive the FK equation for the characteristic function of a general stochastic functional when the random walk is described by the Generalized Master equation of the CTRW. In Sections \ref{OT-interval} and \ref{OT-half-interval} we study the statistical properties (mean, fluctuation, and the PDF) for the occupation time in an interval and for the half occupation time. Finally, in Section \ref{OT-resetting} we analyze how these statistical properties are modified by a Poissonian resetting of the walker's position. We summarize our work with future outlook in Section \ref{Summary}. Some details of the calculations are moved to the Appendices for the clarity of the presentation.   

\section{Non-Markovian random walk}
\label{CTRW-formalism}
We use the Continuous-Time Random Walk (CTRW) framework put forward by Montroll and Weiss to deal with non-Markovian random walks \cite{MoWe65}. We refer to \cite{metzler2000random} for an extensive review of this subject. In the current set-up, we assume that the walker begins at the initial position at time $t=0$ waiting a random time before performing the first jump to a new position. It then waits again for a time interval before proceeding to the next jump. Jump lengths and waiting times are independent and identically distributed random variables according to the PDFs $\Psi(x)$ and $\varphi(t)$, respectively. When the waiting time PDF is exponential then the random walk is Markovian, otherwise, it is non-Markovian. We first define the Laplace and Fourier transforms of the waiting time- and jump length- distributions as $\varphi(s)$ and $\Psi(k)$ respectively  
\[
\varphi(s)=\mathcal{L}_{t\to s}\left[\varphi(t)\right]=\int_{0}^{\infty}e^{-st}\varphi(t)dt,
\]
and 
\begin{eqnarray}
    \Psi(k)=\mathcal{F}_{x\to k}[\Psi(x)]=\int_{-\infty}^{\infty}e^{ikx}\Psi(x)dx.
    \label{ft}
\end{eqnarray}
The propagator $P(x,t|x_{0})$ that describes the position of the walker at time $t$ given that it had started at the location $x_{0}$ is given by the famous Montroll-Weiss equation \cite{MoWe65}, which in Fourier-Laplace space reads 
\begin{equation}
P(k,s|x_{0})=\frac{e^{ikx_{0}}\left[1-\varphi(s)\right]}{s\left[1-\varphi(s)\Psi(k)\right]}.\label{eq:mw}
\end{equation}
Rearranging Eq. (\ref{eq:mw}) in the form 
\begin{equation}
s\left[\frac{1}{\varphi(s)}-\Psi(k)\right]P(k,s|x_{0})=e^{ikx_{0}}\left[\frac{1}{\varphi(s)}-1\right],\label{eq:mw2}
\end{equation}
we can rewrite the expression as 
\begin{equation}
sP(k,s|x_{0})-e^{ikx_{0}}=K(s)\left[\Psi(k)-1\right]P(k,s|x_{0}),\label{eq:mw21}
\end{equation}
where we have introduced the memory kernel 
\begin{equation}
K(s)=\frac{s\varphi(s)}{1-\varphi(s)},\label{eq:mk}
\end{equation}
where $K(t)$ is the memory kernel in real time. 
Note that when the memory kernel $K(t)$ is a Dirac delta function then the random walk is said to be memoryless or Markovian. Otherwise, the random walk is considered to be non-Markovian. If the PDF of jumps $\Psi (x)$ has finite moments we can consider the continuum limit in space and expand it in the Fourier space up to the second order in $k$. Expanding the exponential function in Eq. \eqref{ft} in power series of $k$ one has $\Psi(k)\simeq 1-M_2  k^{2}/2$ where the mean square jump distance (second moment) is $M_2 =\int_{-\infty}^{\infty}x^{2}\Psi(x)dx$. We denote by $\sigma$ the characteristic dispersal distance defined as $\sigma = \sqrt{M_2}$.
For example, if the walker jumps a distance $l$ to the right or left with probability 1/2, then $\Psi (k)=\cos (kl)\simeq 1-k^2l^2/2+...$, so that in this case $\sigma =l$. On setting $\Psi(k)\simeq 1-  (\sigma k)^{2}/2$ in Eq. (\ref{eq:mw21}) and inverting in Fourier-Laplace we obtain the Generalized Master Equation (GME) for random walks 
\begin{equation}
\frac{\partial P(x,t|x_{0})}{\partial t}=\frac{\sigma^{2}}{2}\int_{0}^{t}K(t-t')\frac{\partial^{2}P(x,t'|x_{0})}{\partial x^{2}}dt'.\label{eq:me}
\end{equation}
It is easy to find the Mean Square Displacement (MSD) by making use of the propagator in the Fourier-Laplace space
\begin{eqnarray}
    \left\langle x^{2}(s)\right\rangle =-\left(\frac{\partial^{2}P(k,s|x_{0})}{\partial k^{2}}\right)_{k=0}.
    \label{msd1}
\end{eqnarray}
Using \eqref{eq:me} and \eqref{msd1}, the MSD can be obtained in terms of the waiting time PDF as
\begin{eqnarray}
    \left\langle x^{2}(t)\right\rangle =\sigma^{2}\mathcal{L}_{s\to t}^{-1}\left[\frac{\varphi(s)}{s\left(1-\varphi(s)\right)}\right],
    \label{msd2}
\end{eqnarray}
where we have assumed that the walker starts from the origin. For a Markovian random walk the waiting time PDF is $\varphi(t)=e^{-t/\left\langle \tau\right\rangle}/\left\langle \tau\right\rangle$ and thus, from \eqref{msd2} the MSD is $\left\langle x^{2}(t)\right\rangle =\sigma ^2t/\left\langle \tau\right\rangle$. Here, we have assumed that $\left\langle \tau\right\rangle = \int_0^\infty t\varphi (t)dt$ is the first moment of the waiting time PDF, i.e., the mean waiting time.
If the waiting time PDF has all the moments finite then, for small $s$, one can do the following expansion
\begin{eqnarray}
    \varphi(s)&=&\int_{0}^{\infty}\left(1-st+\frac{1}{2!}s^{2}t^{2}+...\right)\varphi(t)dt\nonumber\\
    &=&1-\left\langle \tau\right\rangle s+O(s^{2}),
    \label{fiapfm}
\end{eqnarray}
and thus the memory kernel is $K(s)\simeq 1/\left\langle \tau\right\rangle $ (i.e., $K(t)\simeq \delta (t)/\left\langle \tau\right\rangle$).
%where $\left\langle \tau\right\rangle = \int_0^\infty t\varphi (t)dt$ is the first moment of the waiting time PDF, i.e., the mean waiting time.  
In this case Eq. \eqref{eq:me} reduces to the diffusion equation and inserting Eq. \eqref{fiapfm} into  Eq. \eqref{msd2} we obtain the same MSD in the long time limit as for the exponential waiting time PDF. This means that the MSD, for any random walk with a waiting time PDF with finite moments, converges to $\left\langle x^{2}(t)\right\rangle =\sigma ^2t/\left\langle \tau\right\rangle=2Dt$ with $D=\sigma^2/2\left\langle \tau\right\rangle$, which corresponds to the MSD of normal diffusion. There are many particular examples of waiting time PDFs with finite moments, besides the exponential case \cite{So02}. An interesting example is the sharp waiting time $\varphi(t)=\delta (t-\tau)$ which would correspond to the discrete-time random walk. 

We now turn our attention to the anomalous case.
To account for the anomalous diffusion of the walker's movement we first consider the waiting time PDF 
with power-law tails. In this case, in the long time limit 
$$
 \varphi(t)\sim \frac{1}{t^{1+\alpha}}.
$$
As we show below, the full form of $\varphi (t)$ is of importance for the occupation time statistics. We consider two cases of power-law waiting-times PDFs. The first example is the fat-tailed PDF \cite{MeKl01}

\begin{eqnarray}
 \varphi(t)=\left\{ \begin{array}{ll}
0, & t<t_{0}\\
\frac{\alpha t_{0}^{\alpha}}{t^{1+\alpha}}, & t>t_{0},
\end{array}\right.
\label{fit}
\end{eqnarray}
where the index $0<\alpha<2$ and $t_0$ is a timescale. Note that for $0<\alpha<1$ the waiting time PDF lacks of finite moments while for $1<\alpha<2$ only the first moment is finite.  The Laplace transform of Eq. \eqref{fit} is given by
\begin{eqnarray}
    \varphi(s)=\alpha e^{-st_{0}}U(1,1-\alpha,st_{0}),
    \label{ltft}
\end{eqnarray}
where $U(a,b;z)$ is the Tricomi confluent hypergeometric functions of the second kind \cite{ab64}. Using the Tauberian theorem \cite{Feller} in Eq. \eqref{ltft}, we find the following expansion in Laplace space for small $s$
\begin{eqnarray}
   \varphi(s)\simeq\left\{ \begin{array}{ll}
1-b_{\alpha}s^{\alpha}+..., & 0<\alpha<1\\
1-\left\langle \tau\right\rangle s+b_{\alpha}s^{\alpha}+..., & 1<\alpha<2
\end{array}\right. 
\label{FI}
\end{eqnarray}
where
$b_{\alpha}=t_{0}^{\alpha}|\Gamma(1-\alpha)|$ and $\left\langle \tau\right\rangle =\alpha t_0/(\alpha -1)$.
The second example is the waiting time PDF of Mittag-Leffler kind
\begin{eqnarray}
    \varphi(t)=\frac{t^{\alpha-1}}{\tau^{*\alpha}}E_{\alpha,\alpha}\left(-\frac{t^{\alpha}}{\tau^{*\alpha}}\right),
    \label{FII}
\end{eqnarray}
with $0<\alpha <1$ and $E_{\alpha,\beta}(\cdot)$ is the two-parametric Mittag-Leffler function \cite{Go20}. The PDF takes the following form in Laplace space
\begin{eqnarray}
    \varphi(s)=\frac{1}{1+(s\tau^*)^{\alpha}},
    \label{MLlt}
\end{eqnarray}
which will be used later in the paper. 

\section{Feynman-Kac equation}
\label{FK-formalism}
The Feynman-Kac formulation has been instrumental to describe the statistical properties of path functionals along a stochastic trajectory. Here, we delineate a few steps of this formalism relevant to our problem. We start by considering the stochastic functional 
\begin{equation}
    Z(t|x_0)=\int _0 ^t U[x(t')]dt',
    \label{Z}
\end{equation}
where $U[x(t')]$ is a positive function of the stochastic trajectory $\{x(t'); 0\leq t'\leq t\}$ of the random walker starting from $x_0$ at $t=0$. Clearly, $Z$ is a stochastic functional that varies between the realizations, and we denote its as $Q(Z,t|x_0)$. To make further progress, it is quite useful to  represent the characteristic function or the moment generating function (MGF) of the functional $Z$ as the double Laplace transform of its PDF $Q(Z,t|x_0)$ which we do next. First, we take the following Laplace transform 
\begin{align}
     Q(p,t|x_0)&=\mathcal{L}_{Z\to p}[Q(Z,t|x_0)]\nonumber\\
  &=\int_0 ^\infty dZ e^{-pZ}Q(Z,t|x_0)  ,
\end{align}
which satisfies the following backward 
master equation (see Appendix \ref{Appendix-A} for the detailed derivation)
\begin{eqnarray}
       \frac{\partial Q(p,t|x_{0})}{\partial t}&=&\frac{\sigma^{2}}{2}\int_{0}^{t}K(t')e^{-pt'U(x_{0})}\frac{\partial^{2}Q(p,t-t'|x_{0})}{\partial x_{0}^{2}}dt'\nonumber\\
       &-&pU(x_{0})Q(p,t|x_{0}).
    \label{FK} 
\end{eqnarray}
Following this, we also define another Laplace transform conjugate to time as
\begin{eqnarray}
  Q(p,s|x_0)&=&\mathcal{L}_{t\to s}[Q(p,t|x_0)]\nonumber\\
  &=&\int_0 ^\infty dt~ e^{-st}Q(p,t|x_0)  ,
  \label{Qps}
\end{eqnarray}
which satisfies the following equation (see Appendix \ref{Appendix-A})
\begin{eqnarray}
sQ(p,s|x_{0})-1&=&\frac{\sigma^{2}}{2}
K(s+pU(x_{0}))
\frac{\partial^{2}Q(p,s|x_{0})}{\partial x_{0}^{2}}\nonumber\\
&-&pU(x_{0})Q(p,s|x_{0}).\label{eq:main-bectrw4}
\end{eqnarray}
In essence, the above is the backward Feynman-Kac equation for a stochastic functional of a non-Markovian random walker. 

%For clarity of the presentation, we have consigned the details of the derivation to Appendix A. 

The moments of $Z (t|x_0)$ can be obtained from the derivatives of $Q(p,s|x_0)$  in a systematic manner. To do this, we first define the Laplace transform of the moments 
\begin{align}
    \left\langle Z(s|x_{0})^{n}\right\rangle &\equiv\mathcal{L}_{t\to s}[\left\langle Z(t|x_{0})^{n}\right\rangle ] \nonumber \\
   & = \int_0^\infty dt ~e^{-st} \int_{0}^{\infty}Z(t|x_0)^{n}Q(Z,t|x_{0})dZ.
\end{align}
By noting the steps below
\begin{align}
    \left\langle Z(s|x_{0})^{n}\right\rangle = 
    \int_0^\infty dt ~e^{-st} \int_{0}^{\infty} \left[(-1)^n \frac{\partial^n}{\partial p^n} e^{-pZ} \right]_{p=0}
    Q(Z,t|x_{0})dZ,
\end{align}
it is then straightforward to show that the moments are represented in terms of the MGF in the following way
\begin{align}
    \left\langle Z(s|x_0)^{n}\right\rangle =(-1)^{n} \frac{\partial^{n}Q(p,s|x_0)}{\partial p^{n}}\bigg|_{p=0}. 
    \label{moments}
\end{align}
In what follows, we consider two specific stochastic functionals that are interesting in analyzing the properties of the search processes: the occupation time in an interval and the occupation time along the half-positive real line. For each case, we solve for the MGF from Eq. \eqref{eq:main-bectrw4} using suitable boundary conditions, and then use Eq. \eqref{moments} to derive the moments which we analyze in detail.

% with respect to $p$. From the definition of the Laplace transform of the PDF with respect to $Z$ we find
% \begin{eqnarray}
%     Q(p,t|x_{0})&=&\mathcal{L}_{Z\to p}[Q(Z,t|x_{0})]=\int_{0}^{\infty}dZe^{-pZ}Q(Z,t|x_{0})\nonumber\\
%     &=&\sum_{n=0}^{\infty}\frac{(-p)^{n}}{n!}\int_{0}^{\infty}Z^{n}Q(Z,t|x_{0})dZ\nonumber\\
%     &=&\sum_{n=0}^{\infty}\frac{(-p)^{n}}{n!}\left\langle Z^{n}(t|x_0)\right\rangle .
%     \label{expa}
% \end{eqnarray}
% On the other hand, expanding $Q(p,t|x_{0})$ in Taylor series around $p=0$
% $$
% Q(p,t|x_{0})=\sum_{n=0}^{\infty}\frac{p^{n}}{n!}\left(\frac{\partial^{n}Q(p,t|x_{0})}{\partial p^{n}}\right)_{p=0}
% $$
% and comparing with \eqref{expa} one obtains
% \begin{eqnarray}
%     \left\langle Z(s|x_0)^{n}\right\rangle =(-1)^{n}\left(\frac{\partial^{n}Q(p,s|x_0)}{\partial p^{n}}\right)_{p=0},
%     \label{moments}
% \end{eqnarray}
% where we use the notation . 

\section{Occupation time in an interval}
\label{OT-interval}
We first consider the occupation time $T_{a} (t|x_0)$ of the random walker, starting initially from $x_0$, in an interval $[-a,a]$ up to an observation time $t$. In this case, the function $U(x_0)=\theta (-a<x_0<a)$ with $\theta (\cdot)$ being the Heaviside step function. The corresponding Feynman-Kac equation \eqref{eq:main-bectrw4} needs to be solved with suitable boundary conditions. For instance, if the starting point is at infinity i.e., $x_{0}\rightarrow \pm\infty$ the walker will never reach the interval so that $P(T_a,t|x_{0}\rightarrow \pm\infty)=\delta (T_a)$, i.e., $Q(p,s|x_{0}\rightarrow \pm\infty)=1/s$. Skipping details from Appendix \ref{Appendix-B}, the solution to the characteristic function \eqref{eq:main-bectrw4} in the Laplace space can be found as
\begin{eqnarray}
 Q(p,s)=Q(p,s|x_{0}=0)=\frac{1}{s+p}\left[1+\frac{p/s}{F(s,p)}\right],
 \label{QTA}
\end{eqnarray}
where
\begin{eqnarray}
   \lambda(s)=a\sqrt{\frac{2s}{\sigma^{2}K(s)}},
   \label{lambda}
\end{eqnarray}
and
\begin{align}
F(s,p)=\cosh\left(\lambda(s+p)\right)+\frac{\lambda(s+p)}{\lambda(s)}\sinh\left(\lambda(s+p)\right).
   \label{Fsp}
\end{align}
Here, we have set  $x_0=0$ without any loss of generality. 
%Eq. \eqref{QTA} is derived explicitly in Appendix B. 

\subsection{Moments}
The moments of $T_a$ can be computed using \eqref{moments} such that
\begin{eqnarray}
    \left\langle T_a(s)^{n}\right\rangle =(-1)^{n}\left(\frac{\partial^{n}Q(p,s)}{\partial p^{n}}\right)_{p=0},
    \label{moments1}
\end{eqnarray}
where $Q(p,s)$ is given by Eq. \eqref{QTA}. 
The first moment is therefore given by
\begin{eqnarray}
  \left\langle T_{a}(s)\right\rangle =\frac{1}{s^{2}}\left[1-e^{-\lambda(s)}\right] , 
  \label{m1}
\end{eqnarray}
and the second moment reads
\begin{eqnarray}
   \left\langle T_{a}(s)^{2}\right\rangle &=&\frac{2}{s^{3}}\left[1-e^{-\lambda(s)}\right. \nonumber\\
   &-&\left. \Phi(s)e^{-\lambda(s)}\left(\lambda(s)+\frac{1}{2}-\frac{e^{-2\lambda(s)}}{2}\right)\right] ,
   \label{m2}
\end{eqnarray}
where
\begin{eqnarray}
  \quad\Phi(s)=\frac{1}{2}-\frac{s}{2K(s)}\frac{dK(s)}{ds}.  
  \label{Fi}
\end{eqnarray}
As $\varphi (s)$ is a monotonically increasing function of $s$, one can easily show that $\lambda (s)$ increases monotonically with $s$. Then, $\lambda (s)$ goes to zero as $s\to 0$. Using this argument we can approximate the expressions \eqref{m1} and \eqref{m2} in the long time limit. Expanding the exponential functions of $\lambda (s)$ we find
\begin{eqnarray}
    \left\langle T_{a}(s)\right\rangle \simeq\frac{\lambda(s)}{s^{2}},
    \label{m1l}
\end{eqnarray}
and
\begin{eqnarray}
\left\langle T_{a}(s)^{2}\right\rangle \simeq\frac{2\lambda(s)}{s^{3}}\left[1-\frac{\lambda(s)}{2}-\Phi(s)\left(2-3\lambda(s)\right)\right],
    \label{m2l}
\end{eqnarray}
as $s\to 0$. Expressions \eqref{m1l} and \eqref{m2l} are the long time limit (as $s \to 0$ corresponds to $t \to \infty$) of the mean and mean squared occupation time in an interval in the Laplace space. To proceed further we now consider explicit forms of the waiting time PDF.

First, we consider the case of a waiting time PDF with finite moments (FM) as in Eq. \eqref{fiapfm}. 
Using $K(s) \simeq 1/\left\langle \tau\right\rangle$ from Eq. \eqref{fiapfm} in Eq. \eqref{lambda} yields $\lambda (s)\simeq a\sqrt{2s\left\langle \tau\right\rangle/\sigma^2}$ as $s\to 0$ which can be substituted into Eq. \eqref{m1l} to arrive at the mean occupation time for a diffusing walker
\begin{eqnarray}
    \left\langle T_{a}(t)\right\rangle_{\rm{FM}} \simeq\frac{2a\sqrt{t}}{\sqrt{\pi D}},
    \label{Tabm}
\end{eqnarray}
in the long time limit. Following a similar derivation for the second moment using Eqs. \eqref{m2l} and \eqref{lambda}, we find
\begin{eqnarray}
    \left\langle T_{a}(t)^{2}\right\rangle _{\rm{FM}}\simeq\frac{2a^{2}}{D}t.
    \label{Tabm2}
\end{eqnarray}
Results \eqref{Tabm} and \eqref{Tabm2} were recently derived by other means and reported in Ref. \cite{BaFlMe24}.

\begin{figure}
    \centering
    \includegraphics[width=0.9\linewidth]{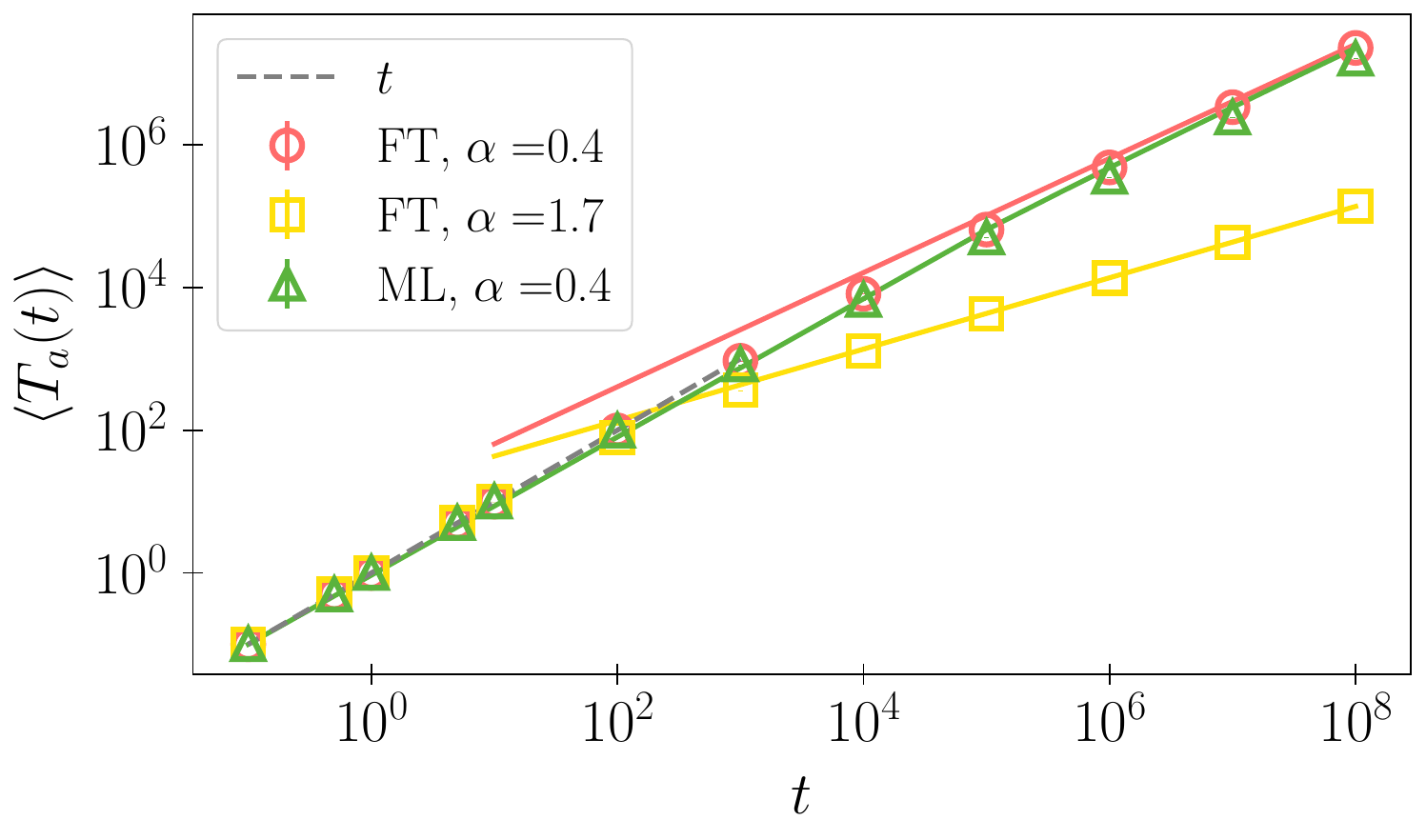}
    \caption{Mean occupation time in an interval. The points are computed with numerical simulations: the red circles and the yellow squares are computed with the FT waiting time PDF given by Eq. \eqref{fit}, and the green triangles are computed with the ML waiting time PDF given by Eq. \eqref{FII}. The solid lines represent theoretical predictions from Eq. \eqref{TA1}  and Eq. \eqref{ta2} respectively. The parameters used are $a=0.055$, $x_0=0$, $\sigma=0.01$, $t_0=1$ for the FT, and $\tau^*=|\Gamma(1-\alpha)|^{1/\alpha}$ for the ML. The number of walkers for the simulations is $N=10^5$. Both axes are in log-log scale.}
    \label{fig:Ta}
\end{figure}

Next, we consider the two examples waiting-time PDFs with power-law tails given by Eqs. \eqref{fit} and \eqref{FII}. In the following we indicate the sub-index FT when the quantity is computed using the fat-tailed waiting time PDF given in Eq. \eqref{fit} while sub-index ML refers to the quantity computed using the Mittag-Leffler PDF in Eq. \eqref{FII}.  For the first case, inserting Eqs. \eqref{eq:mk} and \eqref{FI} in Eq. \eqref{m1l} and inverting the Laplace transform (see Appendix \ref{Appendix-C} for derivation), we find
\begin{eqnarray}
\left\langle T_{a}(t)\right\rangle_{\text{FT}} \simeq\left\{ \begin{array}{ll}
a\sqrt{\frac{2b_{\alpha}}{\sigma^{2}}}\frac{t^{1-\frac{\alpha}{2}}}{\Gamma\left(2-\frac{\alpha}{2}\right)}, & 0<\alpha<1\\
\frac{2a\sqrt{t}}{\sqrt{\pi D}}, & 1<\alpha<2
\end{array}\right. ,
\label{TA1}
\end{eqnarray}
in the $t\to \infty$ limit. 
 It is worth noticing that for a fat-tailed waiting time PDF with $1<\alpha <2$ the mean occupation time is the same as for the case of normal diffusion given by Eq. \eqref{Tabm}.

For the second case, it is possible to obtain an exact solution for the mean occupation time in an interval. Combining Eqs. \eqref{eq:mk}, \eqref{MLlt} and \eqref{lambda}, Eq.  \eqref{m1} turns into
\begin{eqnarray}
    \left\langle T_{a}(s)\right\rangle_\text{ML} =\frac{1}{s^{2}}\left[1-e^{-\frac{a\sqrt{2}}{\sigma}(s\tau^*)^{\alpha/2}}\right],
    \label{tas}
\end{eqnarray}
which can be Laplace inverted to find the temporal behavior in terms of the one-sided L\'evy function $l_{\alpha/2}(t)$ of index $\alpha/2$ 
\begin{eqnarray}
\left\langle T_{a}(t)\right\rangle _{\text{ML}}=\int_{0}^{t}\left\{ 1-L_{\frac{\alpha}{2}}\left[\left(\frac{\sigma}{a\sqrt{2}}\right)^{\frac{2}{\alpha}}\frac{t'}{\tau^{*}}\right]\right\} dt',
\label{ta2}
\end{eqnarray}
where  $L_\alpha(t)$ is the one-sided stable cumulative distribution $L_{\alpha}(t)=\int_{0}^{t}l_{\alpha}(t')dt'$ where the one-sided L\'evy function $l_{\alpha}(t)$ is defined by 
%and the  one-sided L\'evy function is such that its Laplace transform, i.e., 
its characteristic function given by
\begin{eqnarray}
 \int_{0}^{\infty}l_{\alpha}(t)e^{-st}dt=e^{-s^{\alpha}},  \label{os}
\end{eqnarray}
see Ref. \cite{Ba01} and reference therein for more mathematical details on this
function.

In Figure \ref{fig:Ta} we plot the mean occupation time found in Eqs. \eqref{TA1} and \eqref{ta2} together with numerical simulations in a log-log plot. In the large time limit ($s\tau^* \ll 1$) Eq. \eqref{tas} reduces to $\left\langle T_{a}(s)\right\rangle_{\text{ML}}\simeq a\sqrt{2}{\tau^*}^{\alpha/2}s^{-2+\alpha/2}/\sigma$ which back to real time is exactly the result found in Eq. \eqref{TA1} for $0<\alpha<1$ but replacing $b_\alpha$ by ${\tau^*}^{\alpha}$. This can be seen in Figure \ref{fig:Ta}: the red and green lines converge to each other for long times. Note also that in the long-time limit we can keep the first term of the series expansion in Eq. \eqref{ta2} obtaining the same result. Therefore, the mean occupation time in an interval found from the waiting times PDF \eqref{fit} and \eqref{FII} is the same in the long-time limit. This means that the long-time behaviour of the mean occupation time in an interval is governed only by the tails of the waiting time PDF.

Next, we compute the second moment of the occupation time in an interval.  Inserting Eq. \eqref{FI} in \eqref{m2l} we obtain (see Appendix \ref{Appendix-C} for derivation), in the long time limit

\begin{eqnarray}
\left\langle T_{a}(t)^{2}\right\rangle _{\text{FT}}\simeq\left\{ \begin{array}{ll}
\frac{2a(1-\alpha)\sqrt{2b_{\alpha}/\sigma^{2}}}{\Gamma(3-\alpha/2)}t^{2-\frac{\alpha}{2}}, & 0<\alpha<1\\
\frac{2a^{2}}{D}t+\frac{2a(\alpha-1)b_{\alpha}}{\Gamma\left(\frac{7}{2}-\alpha\right)\left\langle \tau\right\rangle \sqrt{D}}t^{\frac{5}{2}-\alpha}, & 1<\alpha<2.
\end{array}\right.
\label{TA2}
\end{eqnarray}

\begin{figure}
    \centering
    \includegraphics[width=0.9\linewidth]{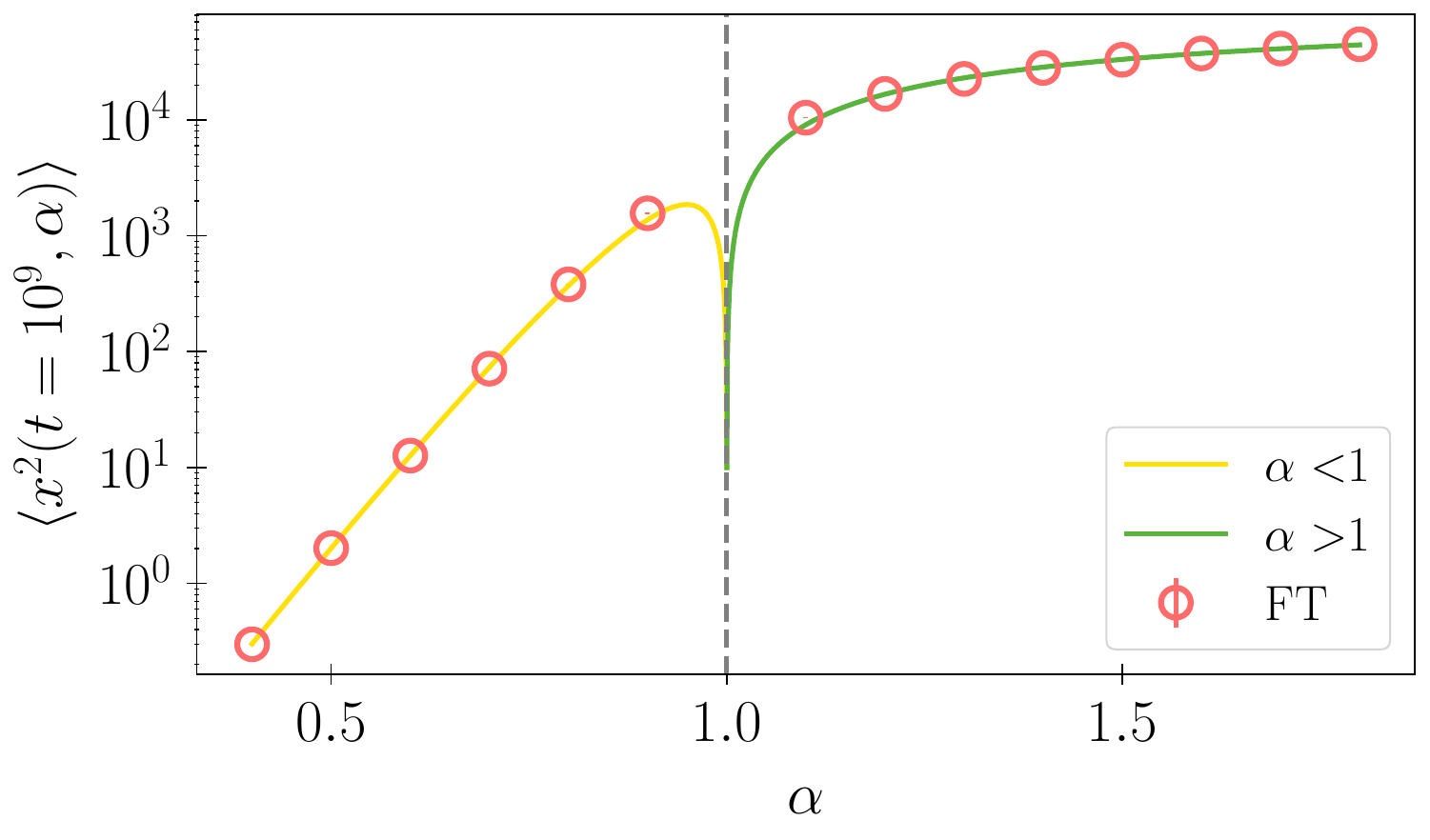}
    \includegraphics[width=0.9\linewidth]{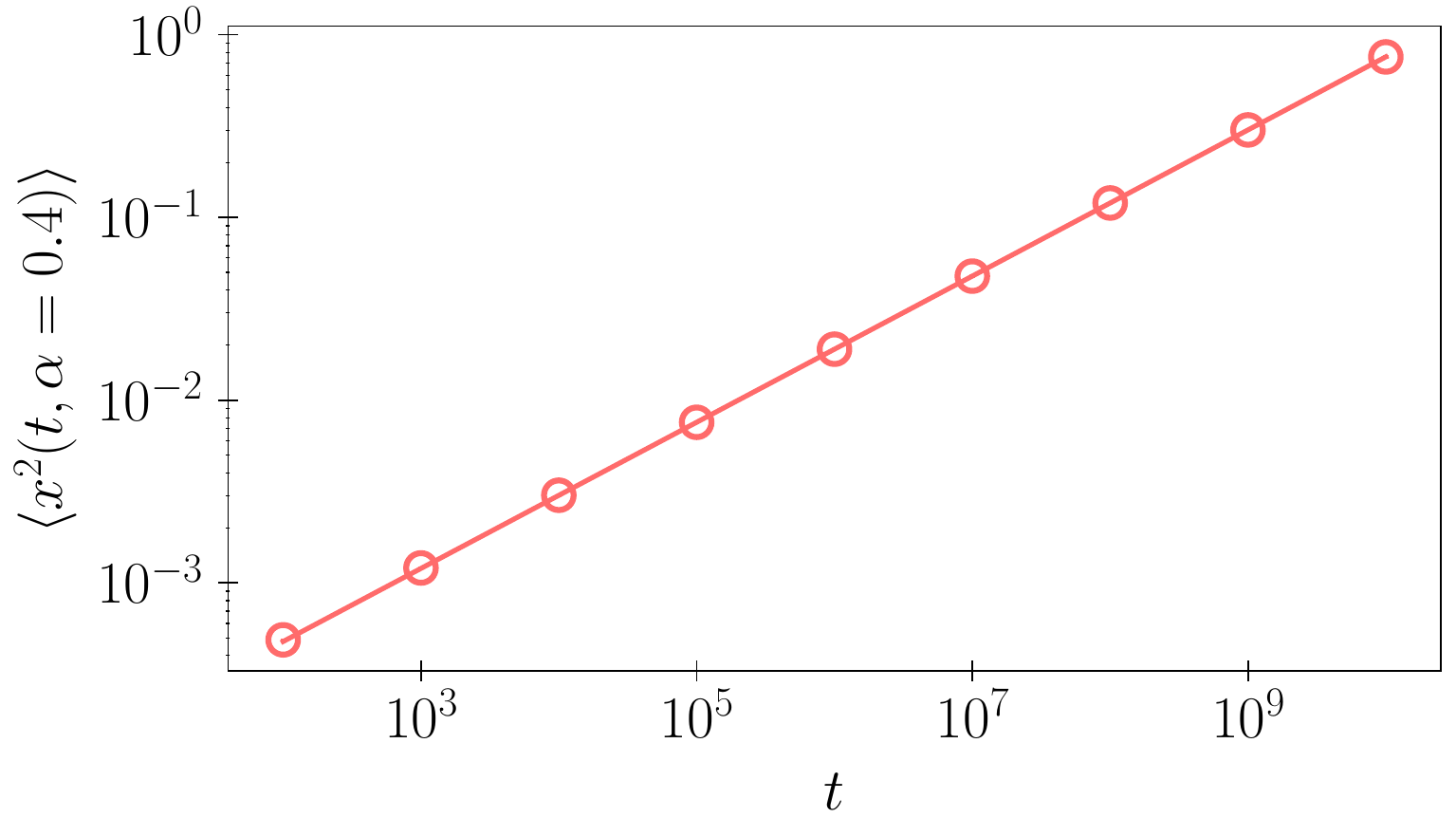}
    \caption{The mean squared displacement (MSD). The points are computed with numerical simulations with the waiting time PDF, Eq. \eqref{fit}, and the lines represent Eq. \eqref{eq:x2}. In the top panel, we fix the maximum simulation time to $t=10^9$ and compute $\langle x^2 \rangle$ for different values of $\alpha$; the dotted vertical line represents the transition value $\alpha=1$. On the other hand, in the bottom panel, we fix $\alpha=0.4$ and show the time evolution of $\langle x^2(t) \rangle$. The parameters used are $a=0.055$, $x_0=0$, $\sigma=0.01$, and $t_0=1$. The number of walkers for the simulations is $N=10^7$. The top panel is plotted in log-lin scale, and the bottom one is in log-log scale.}
    \label{fig:x2}
\end{figure}

\begin{figure}
    \centering
    \includegraphics[width=0.9\linewidth]{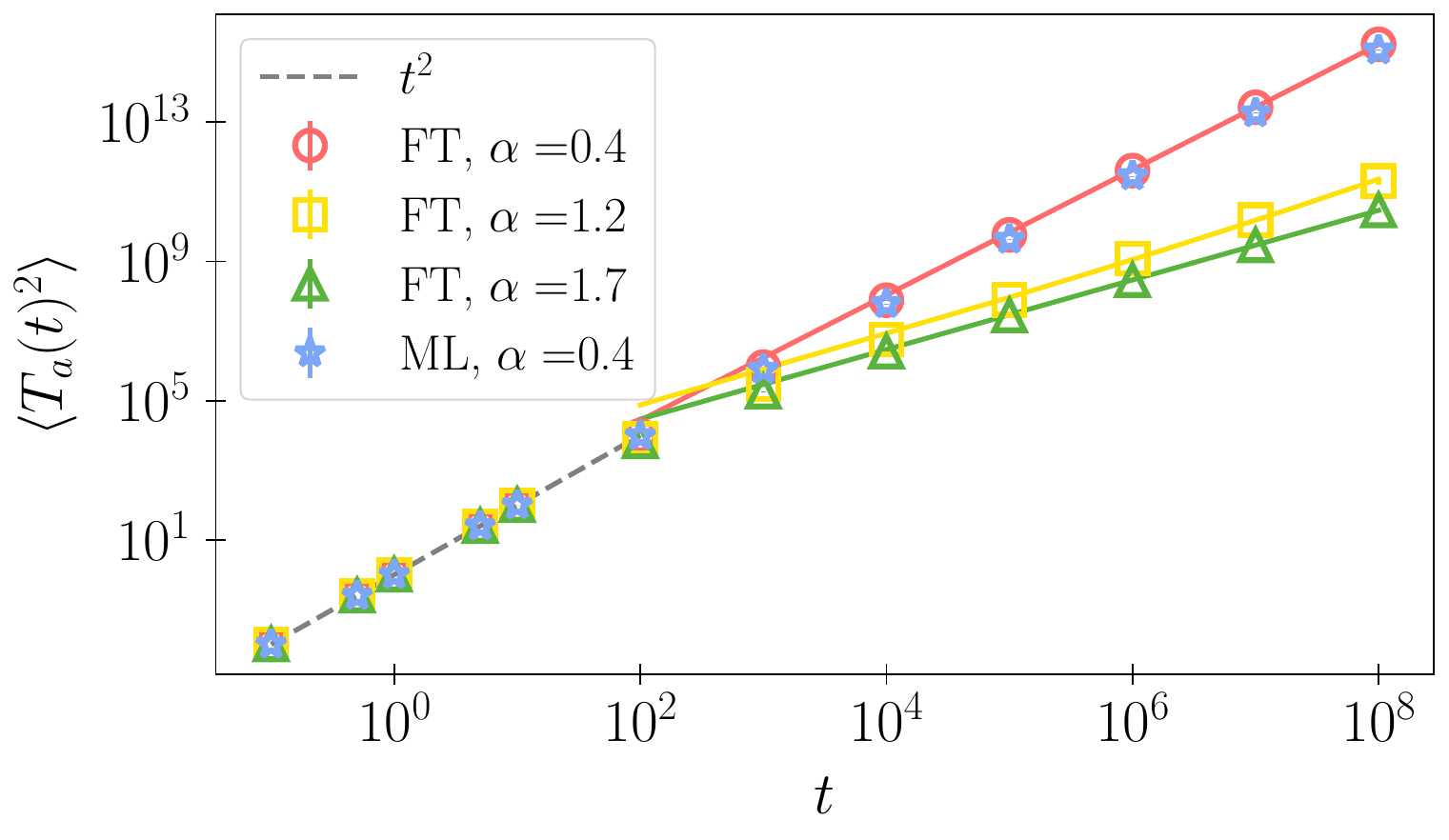}
    \caption{Second moment of the mean occupation time in an interval as a function $t$. The red circles, yellow squares, and green triangles are computed using the FT waiting time PDF, Eq. \eqref{fit}, while the blue stars are with the ML waiting time PDF, Eq. \eqref{FII}. The solid lines represent theoretical predictions from Eq. \eqref{T2a1}. The parameters used here are $a=0.055$, $x_0=0$, $\sigma=0.01$, $t_0=1$ for the power-law data, and $\tau^*=|\Gamma(1-\alpha)|^{1/\alpha}$ for the Mittag-Leffler. The number of walkers per simulation is $N=10^5$. Both axes are in log-log scale.}
    \label{fig:Ta2b}
\end{figure}

It is noteworthy to relate the results for the mean in Eq. \eqref{TA1} and the fluctuations in Eq. \eqref{TA2} for the occupation time to the movement regime of the walker, i.e., with the MSD. To see this connection, we substitute the form of $\varphi(s)$ from \eqref{FI} into Eq. \eqref{msd2}
\begin{eqnarray}
\left\langle x^{2}(t)\right\rangle_{\rm{FT}} \simeq\left\{ \begin{array}{ll}
\frac{\sigma^{2}(t/t_{0})^{\alpha}}{|\Gamma(1-\alpha)|\Gamma\left(1+\alpha\right)}, & 0<\alpha<1\\
\frac{\sigma^{2}}{\left\langle \tau\right\rangle }t, & 1<\alpha<2.
\end{array}\right.
 \label{eq:x2}
\end{eqnarray}
For $0<\alpha<1$ the walker moves subdiffusively while for $1<\alpha<2$ it moves diffusively. In Figure \ref{fig:x2} we compare the previous result for the MSD with numerical simulations. In the upper plot we fix $t=10^9$ and plot the MSD for different values of $\alpha$ ranging from 0 to 2. The value $\alpha=1$ is not included in Eq. \eqref{eq:x2} so that the solid line does not hold in the vicinity of $\alpha=1$. In the lower panel, we plot the MSD vs time in a log-log plot for $\alpha=0.4$. Since for $1<\alpha<2$ the walker moves as a Brownian particle one expects that the mean occupation time and the mean square occupation time are given by Eqs. \eqref{Tabm} and \eqref{Tabm2} respectively. The mean occupation time given in Eq. \eqref{TA1} for $1<\alpha<2$ coincides with Eq. \eqref{Tabm} in the large time limit. However, the mean square occupation time given by Eq. \eqref{TA2} for $1<\alpha<2$ coincides with Eq. \eqref{Tabm2} for $\alpha>3/2$ only. Note that of the two terms on the right side of the equality in \eqref{TA2} the dominant term at long times is the one with the larger exponent, i.e., $\left\langle T_{a}(t)^{2}\right\rangle \sim t^{\max\{1,\frac{5}{2}-\alpha\}}$. Then, for  $1<\alpha<2$ one has $\left\langle T_{a}(t)^{2}\right\rangle \sim t$ if $3/2<\alpha<2$. Summarizing, we can rewrite Eq. \eqref{TA2} in the form 
\begin{align}
    \left\langle T_{a}(t)^{2}\right\rangle _{\text{FT}}\sim\left\{ \begin{array}{ll}
\frac{2a(1-\alpha)}{\Gamma(3-\frac{\alpha}{2})\sqrt{K_{\alpha}}}t^{2-\frac{\alpha}{2}}, & 0<\alpha<1\\
\frac{2a^{2}}{D}t+\frac{2a(\alpha-1)b_{\alpha}}{\Gamma\left(\frac{7}{2}-\alpha\right)\left\langle \tau\right\rangle \sqrt{D}}t^{\frac{5}{2}-\alpha}, & 1<\alpha<\frac{3}{2}\\
\frac{2a^{2}}{D}t, & \frac{3}{2}<\alpha<2
\end{array}\right.
\label{T2a1}
\end{align}
with $K_\alpha = \sigma^2/2b_\alpha$. In Figure \ref{fig:Ta2b} we compare the above result with numerical simulations. Considering the ML distribution from Eq. \eqref{FII} we would get the same result for $\left\langle T_{a}(t)^{2}\right\rangle_\text{ML}$ as in Eq. \eqref{T2a1} for $0<\alpha<1$ in the long time limit \cite{CaBa10,Ba06}. This can be observed in Fig. \ref{fig:Ta2b}. We plot with blue star symbols numerical simulations for $\alpha=0.4$ using the waiting time PDF given in Eq. \eqref{MLlt}. It can be appreciated as these symbols fall on the values (in red circles) obtained for the waiting time \eqref{fit}.

\begin{figure}
    \centering
    \includegraphics[width=\linewidth]{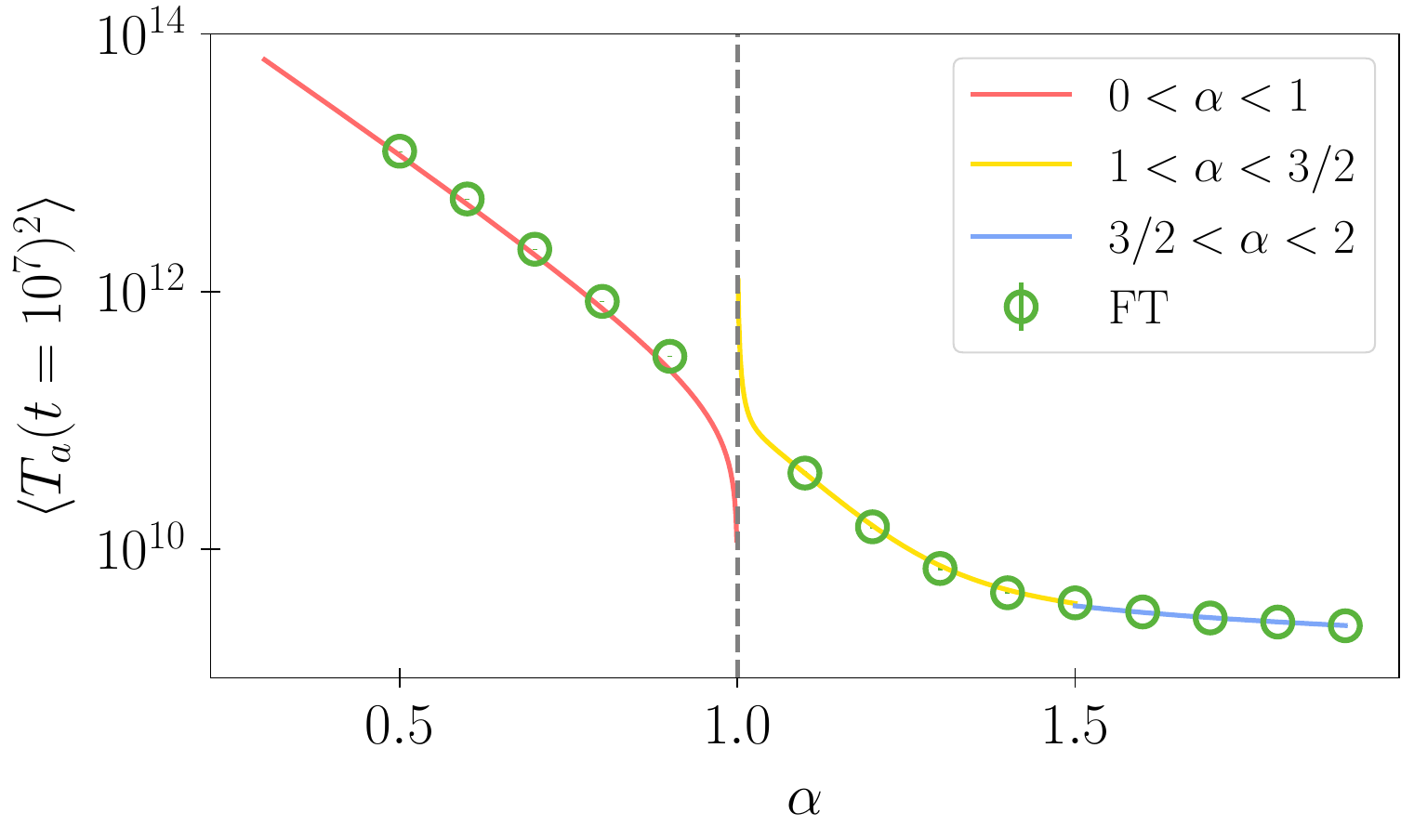}
    \caption{Second moment of the mean occupation time in an interval as a function of $\alpha$. The points are generated from numerical simulations with the waiting time PDF, Eq. \eqref{fit}. The solid lines represent theoretical results from Eq. \eqref{T2a1}. The dashed line marks the transition value $\alpha=1$. The parameters used are $a=0.055$, $x_0=0$, $\sigma=0.01$, and $t_0=1$. The number of walkers for the simulations is $N=10^6$ and the maximum simulation time is $t=10^7$. The plot is presented in log-lin scale.}
    \label{fig:Ta2b_alpha}
\end{figure}

In Figure \ref{fig:Ta2b_alpha} we check the exponents of the temporal dependence of the mean square occupation time given by Eq. \eqref{T2a1} (solid lines) against the numerical simulations (symbols). We note that close to $\alpha=1$ the analytical results diverge. This case should be computed separately but we have not included it here for the sake of simplicity. On the other hand, it is interesting to observe that the short time behavior of the occupation time in an interval is not affected by the underlying random walk, i.e., does not depend on the details of the waiting time PDF. Since the walker is initially at $x_0=0$ (inside the interval), as long as the measurement time is less than the first escape time of the interval $[-a,a]$ the time elapsed by the walker within the interval will be equal to $t$. Therefore, if the measurement time is small enough (smaller than the first escape time of the interval) then $T_a\sim t$ so that $\left\langle T_{a}(t)\right\rangle \sim t$ and $\left\langle T_{a}(t)^2\right\rangle \sim t^2$. This can be checked from Eqs. \eqref{m1} and \eqref{m2}. In the short time limit ($s\to \infty$) Eqs.  \eqref{m1} and \eqref{m2} reduce to $\left\langle T_{a}(s)\right\rangle \simeq s^{-2}$ and $\left\langle T_{a}(s)^2\right\rangle \simeq 2s^{-3}$ since $\lambda (s)\to \infty$ as $s\to \infty$. Finally, inverting the Laplace transform we find $\left\langle T_{a}(t)\right\rangle \simeq t$ and $\left\langle T_{a}(t)^2\right\rangle \simeq t^2$ as $t\to 0$. We have checked this behavior in Figs. \ref{fig:Ta} and \ref{fig:Ta2b}. 

\subsection{PDF}
We now turn our interest to analyze the behavior of $Q(T_{a},t)$ in the long time limit $t\to\infty$. The double Laplace inversion of Eq. \eqref{QTA} is not generically tractable, hence we consider three  specific limits: $T_a\ll t$ with $T_a$ small, $T_a$ comparable to $t$, and $T_a$ close to $t$. These three limits correspond to three regions: left edge, bulk, and right edge. Below, we delve into these regimes.

\subsubsection{Left edge regime}
First we consider  $T_a\ll t$ (i.e., $s\ll p$) with $T_a$ small (corresponding to large $p$). Then, $s+p\simeq p$ and $\lambda(s+p)\simeq \lambda (p)$ and Eq. \eqref{QTA} turns into
\begin{eqnarray}
Q(p,s)\simeq\frac{1}{p}+\frac{s^{-1}}{\cosh(\lambda(p))+\frac{\lambda(p)}{\lambda(s)}\sinh(\lambda(p))}.
    \label{qps0}
\end{eqnarray}
In addition, for large $p$ one has $\cosh(\lambda(p))\simeq 1$ and $\sinh(\lambda(p))\simeq \exp{[\lambda(p)]}/2$. In addition, since $s\ll p$ we have $\lambda(s)\ll\lambda(p)$, so that Eq. \eqref{qps0} reduces to
\begin{eqnarray}
    Q(p,s)\simeq\frac{1}{p}+\frac{2}{s}\frac{\lambda(s)e^{-\lambda(p)}}{\lambda(p)}.
     \label{qps1}
\end{eqnarray}
To proceed further we need to specify explicit expressions for the memory kernel or the waiting time PDF. 
\begin{figure}[h!]
    \centering
    \includegraphics[width=\linewidth]{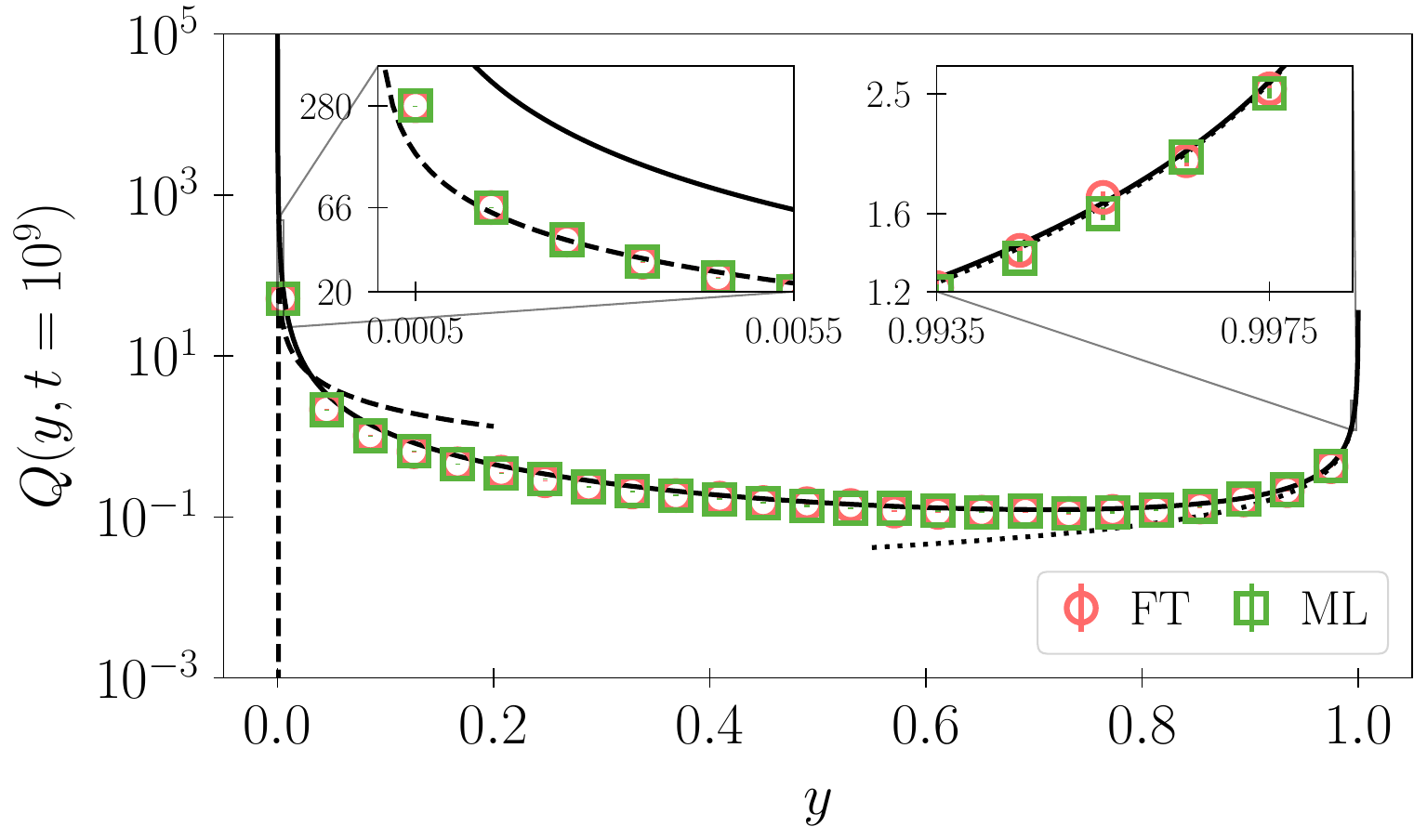}
    \caption{Scaled PDF of occupation time $Q(y,t)$ as a function of $y=T_a/t$ for FT- and ML- waiting times. The points are generated from numerical simulations: the red circles with the waiting time PDF Eq. \eqref{fit} (FT) and the green squares with Eq. \eqref{FII} (ML). Theoretical predictions: The dashed line is computed with Eq. \eqref{QTAT0}, the dotted line with Eq. \eqref{QTAT022}, and the solid line with Eq. \eqref{eq:QTU}. Insets are used to magnify the results from left and right edge regimes. The parameters used are $a=0.055$, $x_0=0$, $\sigma=0.01$, $\alpha=0.4$, $t_0=1$ for the power-law data, and $\tau^*=|\Gamma(1-\alpha)|^{1/\alpha}$ for the Mittag-Leffler. The number of walkers for the numerical simulations is $N=10^6$ and the final time of the simulation is $t=10^9$. The main figure is on a log-lin scale, while the two insets are in lin-lin scale.}
    \label{fig:Qx}
\end{figure}

For the fat-tailed waiting time PDF \eqref{fit} the memory kernel $K(s)$ is given by Eq. \eqref{Kap} for small $s$. Considering $0<\alpha<1$ we find (see Appendix \ref{Appendix-D} for details)
\begin{equation}
Q(T_{a},t)\simeq\frac{2\sin\left(\frac{\pi\alpha}{2}\right)}{\pi}\frac{(T_{a}/t)^{\frac{\alpha}{2}}}{\left(a/\sqrt{K_{\alpha}}\right)^{\frac{2}{\alpha}}}\int_{0}^{1}\frac{l_{\alpha/2}\left[\frac{T_{a}z}{\left(a/\sqrt{K_{\alpha}}\right)^{\frac{2}{\alpha}}}\right]}{(1-z)^{1-\frac{\alpha}{2}}}dz,
\label{QTAT0}
\end{equation}
which holds for $T_{a}\ll(a/\sqrt{K_{\alpha}})^{2/\alpha}$, i.e., in a narrow region very close to $y=0$. In Figure \ref{fig:Qx} we compare numerical simulations with the analytical result in Eq. \eqref{QTAT0} (drawn with a dashed curve). As can be seen from the plot, the agreement is very good (see the left inset).

\begin{figure}
    \centering
    \includegraphics[width=\linewidth]{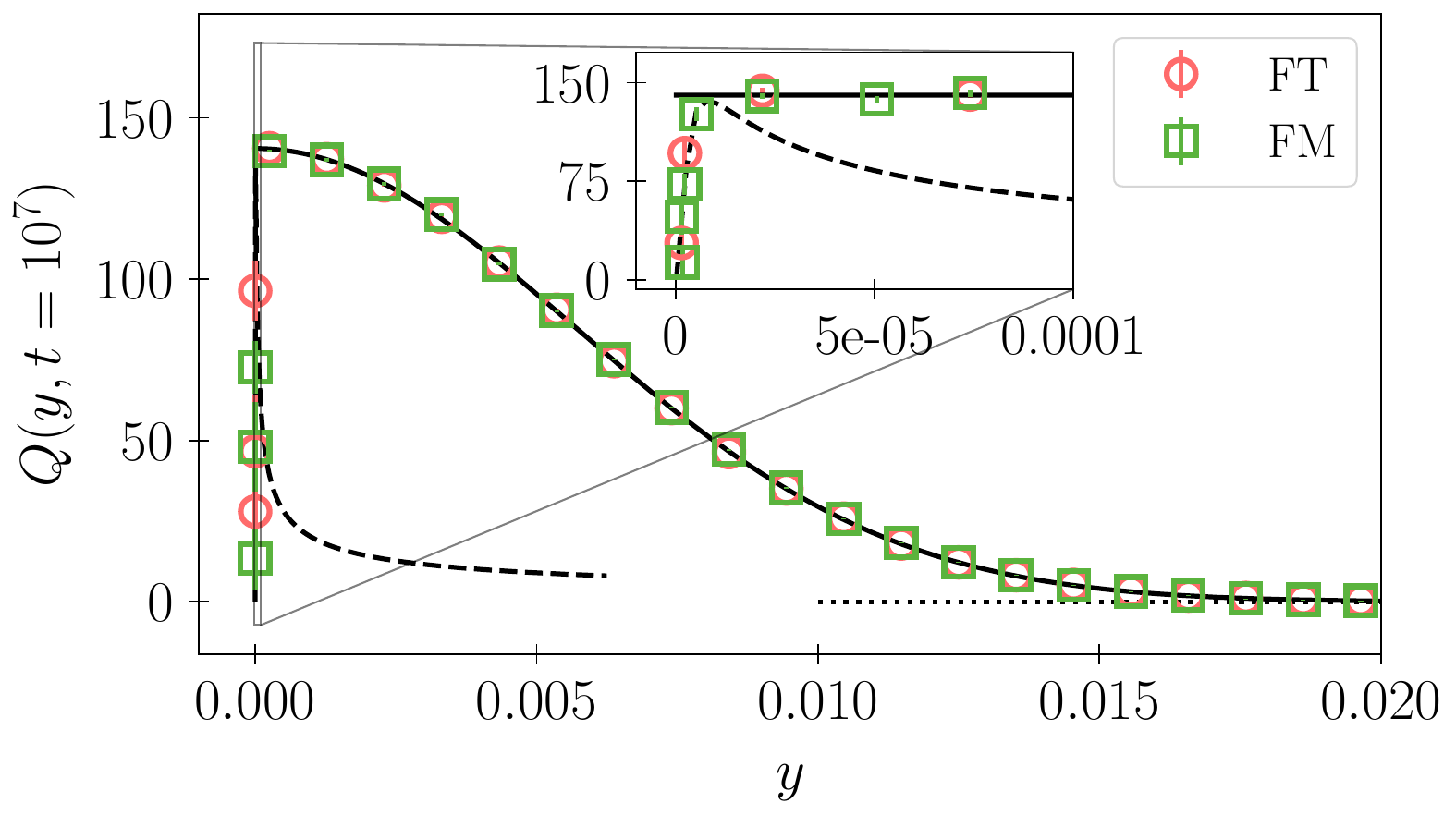}
    \caption{Scaled PDF of occupation time $Q(y,t)$ as a function of $y=T_a/t$ for FM waiting times. The red circles (FT) are for the waiting time PDF in Eq. \eqref{fit} with $\alpha=1.6$; the green squares (FM) are numerically generated from an exponential waiting time PDF $\varphi(t)= e^{-t/\langle \tau \rangle}/\langle \tau \rangle$ with $\langle \tau \rangle = \alpha/(\alpha-1)$. Theoretical predictions: The solid line represents Eq. \eqref{half}, the dashed line Eq. \eqref{QTABM}, and the dotted line represents Eq. \eqref{QREdge_FT}. Inset is used to magnify the results from the left edge. The parameters used for the numerical simulations are $a=0.055$, $x_0=0$, $\sigma=0.01$, $t_0=1$, the number of walkers for the simulation $N=10^6$, and the maximum time $t=10^7$. All axes are in linear scale.}
    \label{fig:Qx_12}
\end{figure}

We consider the case of a waiting time PDF with finite moments. In this case, we have (see Appendix \ref{Appendix-D} for details)
\begin{eqnarray}
Q(T_{a},t)\simeq\frac{2}{\pi\sqrt{tT_{a}}}e^{-\frac{a^{2}}{4DT_{a}}}.
\label{QTABM}
\end{eqnarray}
In Figure \ref{fig:Qx_12} we compare Eq. \eqref{QTABM} (in dashed lines) with the numerical simulations. As can be seen in the inset, the validity of \eqref{QTABM} is restricted to a narrow region very close to $y=0$. The agreement is good until the numerical curve reaches a maximum. 
\\
 
\subsubsection{Right edge regime}
Second, we turn our attention to the large time limit but $T_a\to t$. Define the new random variable $\epsilon = t-T_a$. Let $Q_{\epsilon}(\epsilon,t)$ be the PDF of $\epsilon$ and consider its double Laplace transform $Q_{\epsilon}(p_{\epsilon},s)=\mathcal{L}_{\epsilon\to p_{\epsilon}}\mathcal{L}_{t\to s}\left[Q(\epsilon,t)\right]$. In Laplace space, $Q_{\epsilon}(p_\epsilon,s)=Q(-p_\epsilon,s+p_\epsilon)$ where $Q(-p_\epsilon,s+p_\epsilon)$ is obtained by replacing $p$ by $-p_\epsilon$ and $s$ by $s+p_\epsilon$ in $Q(p,s)$. Thus, 
\begin{eqnarray}
 Q_{\epsilon}(p_{\epsilon},s)=\frac{1}{s}\left[1-\frac{p_{\epsilon}}{(s+p_{\epsilon})F(s+p_{\epsilon},-p_{\epsilon})}\right].
 \label{Qe}
\end{eqnarray}
In the large time limit and $\epsilon\to 0$ (this is, $s\ll p_\epsilon$ ) 
\begin{align}
  Q_{\epsilon}(p_{\epsilon},s)\sim\frac{1}{s}-\frac{\lambda(p_{\epsilon})}{\lambda(p_{\epsilon})+\lambda(s)\tanh(\lambda(s))}\frac{1}{s\cosh(\lambda(s))},  
\end{align}
where we have taken into account Eqs. \eqref{Fsp} and \eqref{Qe}. Taking the large time limit (correspondingly, $s\to 0$) and since $\lambda (s)\to 0$ as $s\to 0$ we have $\cosh(\lambda(s))\simeq 1$ and $\sinh(\lambda(s))\simeq \lambda(s)$, so that
\begin{eqnarray}
   Q_{\epsilon}(p_{\epsilon},s)\sim\frac{1}{s}\frac{\lambda(s)^{2}}{\lambda(p_{\epsilon})+\lambda(s)^{2}}.
    \label{Qpe}
\end{eqnarray}
This is a general expression for the characteristic function of $t-T_a$ in the large time limit and when $t-T_a\ll 1$ which can be found for different expressions of the waiting time PDF.

For example, we can consider the case of a fat-tailed waiting time PDF. We note that in the right edge regime, $T_a$ and $t$ are very large compared to $t_0$ so that $p_\epsilon t_0\ll 1$ and $st_0\ll 1$ and we can make use of Eq. \eqref{lap} from the Appendix \ref{Appendix-C} as expressions for both  $\lambda (s)$ and $\lambda (p_\epsilon)$ for the case of a walker performing rests of power-law distributed duration. Substituting Eq. \eqref{lap} into \eqref{Qpe} we find 
\begin{eqnarray}
 Q(T_{a},t)\simeq\frac{\mathcal{A}_{\alpha}}{t^{\alpha}(t-T_{a})^{1-\frac{\alpha}{2}}},
    \label{QTAT022}
\end{eqnarray}
where $0<\alpha <1$ and
$$
\mathcal{A}_{\alpha}=\frac{a}{\sqrt{K_{\alpha}}\Gamma(\frac{\alpha}{2})\Gamma(1-\alpha)},
$$
after the double Laplace inversion (see details in Appendix \ref{Appendix-E}). If we consider the ML waiting time PDF from Eq. \eqref{FII} we obtain the same results given by Eqs. \eqref{QTAT0} and \eqref{QTAT022} (after replacing $b_\alpha$ by $\tau ^\alpha$), which hold for $0<\alpha <1$. We can express the solution in a scaling form. If we define $y=T_a/t$ then 
\begin{eqnarray}
t^{\alpha/2}Q(y,t)=\frac{\mathcal{A}_{\alpha}}{\left(1-y\right)^{1-\frac{\alpha}{2}}}.
\end{eqnarray}
In Figure \ref{fig:Qx} we observe how the results for $Q(T_{a},t)$ obtained from the numerical simulations performed using the waiting times \eqref{fit} and \eqref{FII} are the same. In addition, expression \eqref{QTAT022} plotted with a dotted curve also agrees very well (see right inset) with simulations in the limit $T_a\to t$.

We now consider the case of a waiting time PDF with finite moments. Then 
$\lambda (s)\simeq a\sqrt{2s\left\langle \tau\right\rangle/\sigma^2}$ as $s\to 0$ and from Eq. \eqref{Qpe}
$$
Q_{\epsilon}(p_{\epsilon},s)\simeq\frac{1}{s}\frac{\frac{as}{\sqrt{D}}}{\sqrt{p_{\epsilon}}+\frac{as}{\sqrt{D}}}.
$$
Performing the double Laplace inversion
\begin{eqnarray}
    Q(T_{a},t)\simeq\sqrt{\frac{D}{4a^{2}\pi}}\frac{t}{(t-T_{a})^{\frac{3}{2}}}e^{-\frac{t^{2}D}{4a^{2}(t-T_{a})}}.
    \label{QREdge_FT}
\end{eqnarray}
In Figure \ref{fig:Qx_12} we plot Eq. \eqref{QREdge_FT} with a dotted line. As can be observed the agreement with the numerical simulations is excellent even in the bulk region.

\subsubsection{Central bulk regime}
Finally, we consider the uniform approximation where $t$ and $T_a$ are both large or $s$ and $p$ small. Then, it is expected that this solution agrees with the limit $T_a\to t$ but does not fit with the limit $T_a\to 0$.  Since both $s$ and $p$ are small, $\lambda(s+p)$ is also small, since $d\lambda(s)/ds>0$. Then, $\cosh\left(\lambda(s+p)\right)\simeq1$ and $\sinh\left(\lambda(s+p)\right)\simeq\lambda(s+p)$ so that
\begin{eqnarray}
    Q(p,s)\simeq\frac{1}{s+p}\left[1+\frac{p}{s}\frac{1}{1+\frac{\lambda(s+p)^{2}}{\lambda(s)}}\right].\label{Qps1}
\end{eqnarray}
Let us be more specific and consider first that the waiting time PDF is FT given in Eq. \eqref{FI} for $0<\alpha<1$ or the ML waiting time \eqref{MLlt}. We have 
\[
Q(p,s)\simeq\frac{1}{s}-a\sqrt{\frac{2b_{\alpha}}{\sigma^{2}}}\frac{p}{s^{1+\frac{\alpha}{2}}(s+p)^{1-\alpha}},
\]
if we assume $O(s)\sim O(p)$.
Performing the double inverse Laplace transform we finally find 
\begin{eqnarray}
Q(T_{a},t)\simeq\mathcal{A}_{\alpha}\frac{\left(t-T_{a}\right)^{\frac{\alpha}{2}}}{T_{a}^{1+\alpha}}\left(2+\frac{T_{a}}{t-T_{a}}\right).
 \label{eq:QTU}
\end{eqnarray}
Note that if $T_a\sim t$, then $2+\frac{T_{a}}{t-T_{a}}\simeq \frac{T_{a}}{t-T_{a}}$ and Eq.\eqref{eq:QTU} reduces to Eq. \eqref{QTAT022}. 
For the fat-tailed PDF with  $1<\alpha<2$, using Eq. \eqref{lap} from the Appendix \ref{Appendix-C}, we get exactly the same result as in Eq. \eqref{half}. Eq. \eqref{eq:QTU} can also be written is scaling form as
\begin{eqnarray}
t^{\alpha/2}Q(y,t)=\mathcal{A}_{\alpha}\frac{2-y}{y^{1+\alpha}\left(1-y\right)^{1-\frac{\alpha}{2}}}.
\end{eqnarray}
In Figure \ref{fig:Qx} we compare Eq. \eqref{eq:QTU} (in solid lines) with numerical simulations. The agreement is very good even in the region where $y$ is close to 1

For a waiting time PDF with finite moments  Eq. \eqref{Qps1} reads
\begin{eqnarray}
Q(p,s)\simeq\frac{1}{s+p}\left[1+\frac{ps^{-1/2}}{s^{1/2}+a\sqrt{\frac{2 \tau }{\sigma^{2}}}(s+p)}\right].  
 \label{QTU2}
\end{eqnarray}
Laplace inverting Eq. \eqref{QTU2} with respect to $p$ and $s$ we find that the PDF for $T_{a}$
follows the half-Gaussian

\begin{eqnarray}
    Q(T_{a},t)\simeq\sqrt{\frac{D}{a^{2}\pi(t-T_{a})}}e^{-\frac{T_{a}^{2}D}{4a^{2}(t-T_{a})}}.
    \label{half}
\end{eqnarray}
In Figure \ref{fig:Qx_12} we compare Eq. \eqref{half} (in solid lines) with numerical simulations. The agreement is again excellent along the central bulk region. 

%We note in passing that the occupation time in the limit $a \to 0$ can be connected to the local time which was discussed in much detail in several other 

\subsection{Ergodic properties}
We move on to study the ergodic properties of the observable $\theta (-a<x(t)<a)$. For brevity, we have revisited the concept of ergodicity and the definition of the ergodicity breaking parameter in the Appendix \ref{Appendix-F}. In here, the time average of the observable $\theta$ is given by $T_a/t$ as
$$
\overline{\theta(-a<x(t)<a)}=\frac{1}{t}\int_{0}^{t}\theta(-a<x(t')<a)dt'=\frac{T_a(t)}{t}.
$$
Then, $\left\langle \overline{\theta(-a<x(t)<a)}\right\rangle =\left\langle T_a(t)\right\rangle /t$ and $\left\langle \overline{\theta(-a<x(t)<a)}^{2}\right\rangle =\left\langle T_a(t)^{2}\right\rangle /t^{2}$. The EB parameter in this case, following \eqref{EB}, reads 
\begin{eqnarray}
    \textrm{EB}=\lim_{t\to\infty}\frac{\left\langle T_a(t)^{2}\right\rangle -\left\langle T_a(t)\right\rangle ^{2}}{\left\langle T_a(t)\right\rangle ^{2}}.
    \label{EBa}
\end{eqnarray}
Inserting Eqs. \eqref{Tabm} and \eqref{Tabm2} for FM together with Eqs. \eqref{TA1} and \eqref{T2a1} for FT into Eq. \eqref{EBa} we find
$$
\textrm{EB}=\left\{ \begin{array}{cc}
\sim t^{\alpha/2}, & 0<\alpha<1\\
\sim t^{\frac{3}{2}-\alpha}, & 1<\alpha<3/2\\
\frac{\pi}{2}-1, & 3/2<\alpha<2.
\end{array}\right.
$$
In consequence, the occupation time in an interval for both Markovian and non-Markovian random walk is non ergodic, regardless of the interval width $a$ by noting that for any ergodic observable, one should have $\textrm{EB}$ identical to zero value \cite{bel2005weak,BaFlMe24}.

\section{Half occupation time}
\label{OT-half-interval}
We consider now the occupation time $T^+ (t|x_0)$ of the walker above the origin (i.e., on the positive half-space) within a time window of size $t$ if it was at $x=x_{0}$ at $t=0$. For this case $U(x_0)=\theta (x_0)$ and the characteristic function has to be solved under the appropriate boundary conditions \cite{Ma05}. If the starting point $x_{0}\rightarrow+\infty$ the walker will stay on the positive side with $x(t)>0$ for all finite $t$ implying $T^{+}(t|x_{0}\rightarrow+\infty)=t$ and accordingly $P(T^+,t|x_{0}\rightarrow+\infty)=\delta(T^+-t),$ i.e., $Q(p,s|x_{0}\rightarrow+\infty)=1/(s+p)$. On the other side, if the starting point $x_{0}\rightarrow-\infty$ the walker will never reach the positive side and it will stay on the negative half-space implying $T^{+}(t|x_{0}\rightarrow-\infty)=0$ and hence $Q(T^+,t|x_{0}\rightarrow-\infty)=\delta(T^+)$, so that $Q(p,s|x_{0}\rightarrow-\infty)=1/s$. The solution for the characteristic function under the above boundary conditions reads (see Appendix \ref{Appendix-G} for a detailed derivation) 
\begin{eqnarray}
    Q(p,s)=\frac{1}{s+p}\left[1+\frac{p}{s+\sqrt{\frac{s(s+p)K(s)}{K(s+p)}}}\right],
    \label{QTP}
\end{eqnarray}
where we have assumed $x_0=0$. This equation allows us to compute the characteristic function for the half occupation time directly from the memory kernel or the waiting time PDF. This means that the half occupation time statistics are governed by the statistics of the resting periods of the random walk. When the waiting time PDF has finite moments, the PDF for the half occupation time $Q(T^+,t)$ admits a limiting distribution in the limit $T^+\to\infty$ and $t\to \infty$ so that it converges to (see Appendix H for the derivation)
\begin{eqnarray}
   Q(y)=\frac{1}{\pi\sqrt{y(1-y)}},
   \label{Qzl}
\end{eqnarray}
where $y=T^+/t$. This is the limiting \textit{arcsine distribution} as one would expect for a diffusive random walker \cite{Ma05,Le40}.

When the waiting time PDF is a power-law we can find a range of behaviors. 
For instance, if it is given by Eq. \eqref{FII}, then the characteristic function $Q(p,s)$ possesses a limiting distribution (the Lamperti PDF \cite{La58,CaBa10}) namely
\begin{align}
Q(y)=\frac{\sin(\frac{\alpha\pi}{2})}{\pi}\frac{y^{\frac{\alpha}{2}-1}(1-y)^{\frac{\alpha}{2}-1}}{y^{\alpha}+(1-y)^{\alpha}+2y^{\frac{\alpha}{2}}(1-y)^{\frac{\alpha}{2}}\cos(\frac{\alpha\pi}{2})},
 \label{lamperti}
\end{align}
which is also the case for Eq. \eqref{fit} with $0<\alpha<1$. However,  for Eq. \eqref{fit} when $1<\alpha<2$ (so that the first moment is finite),  $Q(T,t)$ still follows the \textit{arcsine law} given by Eq. \eqref{Qzl}. The arcsine PDF \eqref{Qzl} is then an \textit{attractor} for the PDF of half occupation time of a random walker having rest periods with finite first moments. This is confirmed in Figure \ref{fig:Qz}. See Appendix \ref{Appendix-H} for the derivation of these limiting distributions.

\begin{figure}[h!]
    \centering
    \includegraphics[width=0.9\linewidth]{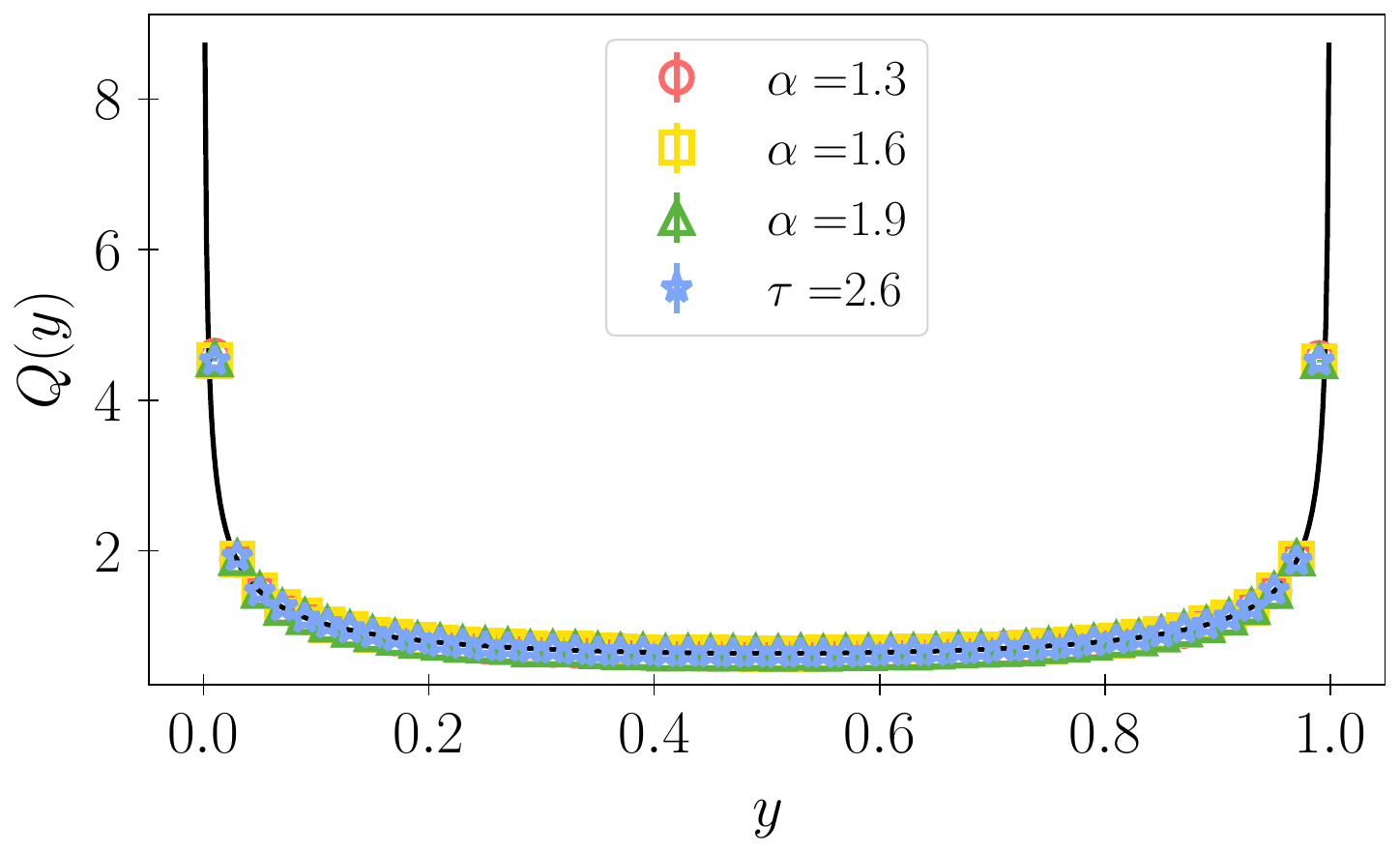}
    \caption{Universality of the \textit{arcsine} distribution. The red circles, yellow squares, and green triangles are computed with the  FT waiting time PDF from Eq. \eqref{fit} for $\alpha=1.3$, $\alpha=1.6$, and $\alpha=1.9$ respectively. The blue stars are computed with an exponential waiting time PDF $\varphi(t)= e^{-t/\langle \tau \rangle}/\langle \tau \rangle$ with $\langle \tau \rangle = 2.6$. The solid line is the arcsine distribution \eqref{Qzl}. The parameters used for the numerical simulations are $x_0=0$, $\sigma=0.01$, $t_0=1$, the number of walkers for the simulation $N=10^6$, and the maximum time $t=10^7$.}
    \label{fig:Qz}
\end{figure}

\subsection{Moments}
\begin{figure}[h!]
    \centering
    \includegraphics[width=0.9\linewidth]{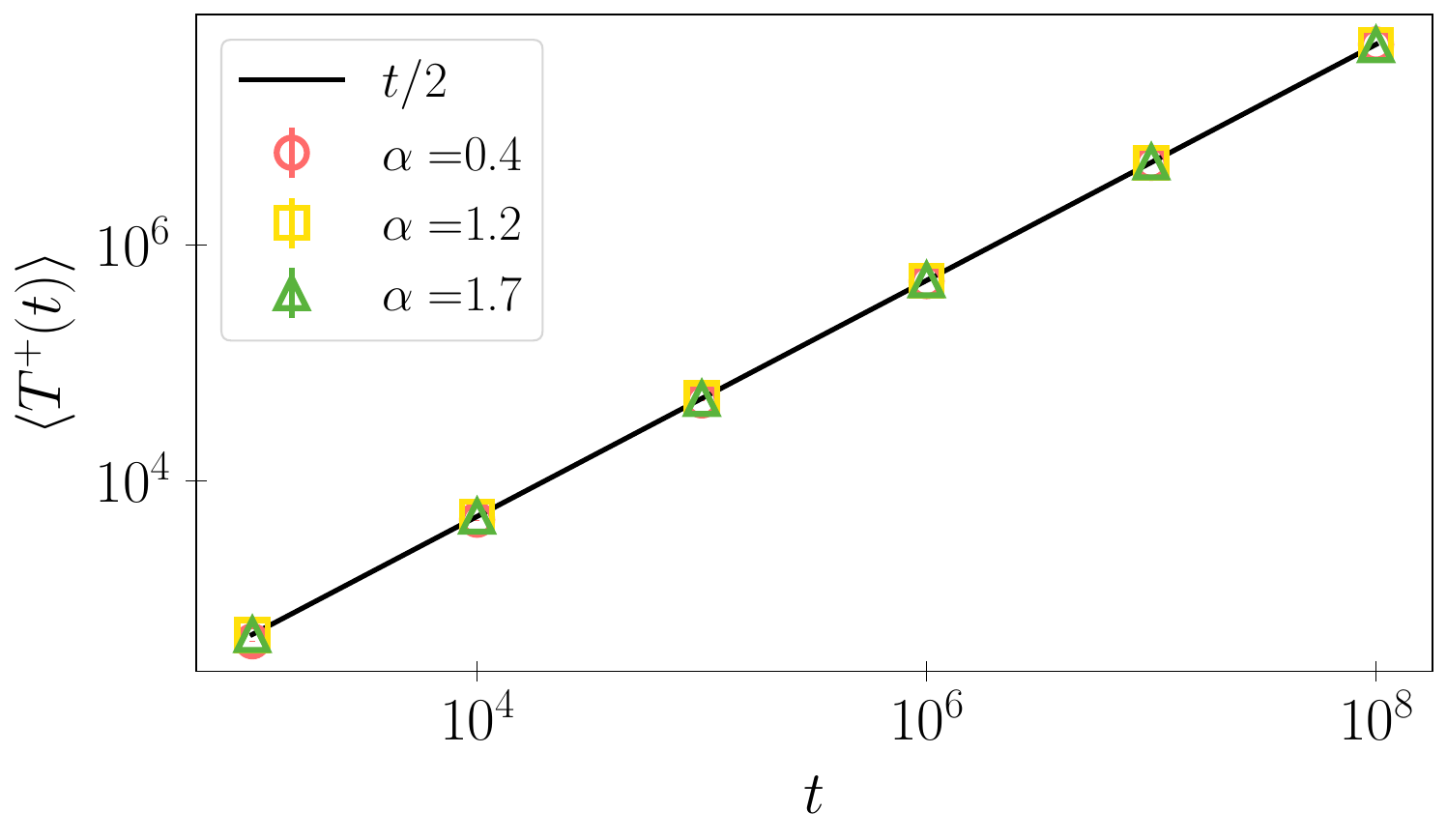}
    \caption{Mean half occupation time. All the data points are generated using the FT waiting time PDF in Eq. \eqref{fit}, the red circles for $\alpha=0.4$, the yellow squares for $\alpha=1.2$, and the green triangles for $\alpha=1.7$. The solid line is a theoretical one and is computed using Eq. \eqref{eq:halfoctime}. The parameters used for the numerical simulations are $x_0=0$, $\sigma=0.01$, $t_0=1$, and the number of walkers for the simulation $N=10^5$. The plot is in log-log scale.}
    \label{fig:Tplus}
\end{figure}

The moments of the half occupation time can be computed using the following relation that was derived earlier
\begin{eqnarray}
    \left\langle T^+(s)^{n}\right\rangle =(-1)^{n}\left(\frac{\partial^{n}Q(p,s)}{\partial p^{n}}\right)_{p=0},
    \label{moments2}
\end{eqnarray}
where $Q(p,s)$ is given by Eq. \eqref{QTP}. Hence, the first moment in Laplace space following Eq. \eqref{moments2} is given by $\left\langle T^{+}(s)\right\rangle =1/2s^{2}$. Inverting, we find
\begin{eqnarray}
\left\langle T^{+}(t)\right\rangle =\frac{t}{2},
\label{eq:halfoctime}
\end{eqnarray}
regardless of the expression for the waiting time PDF, which means that for any isotropic random walk, the mean half occupation time up to a time $t$ is simply $t/2$. This result is compared with numerical simulations in Figure \ref{fig:Tplus} and, as can be appreciated, the mean half occupation time does not depend on the properties of the underlying random walk.

\begin{figure}[h!]
    \centering
    \includegraphics[width=0.9\linewidth]{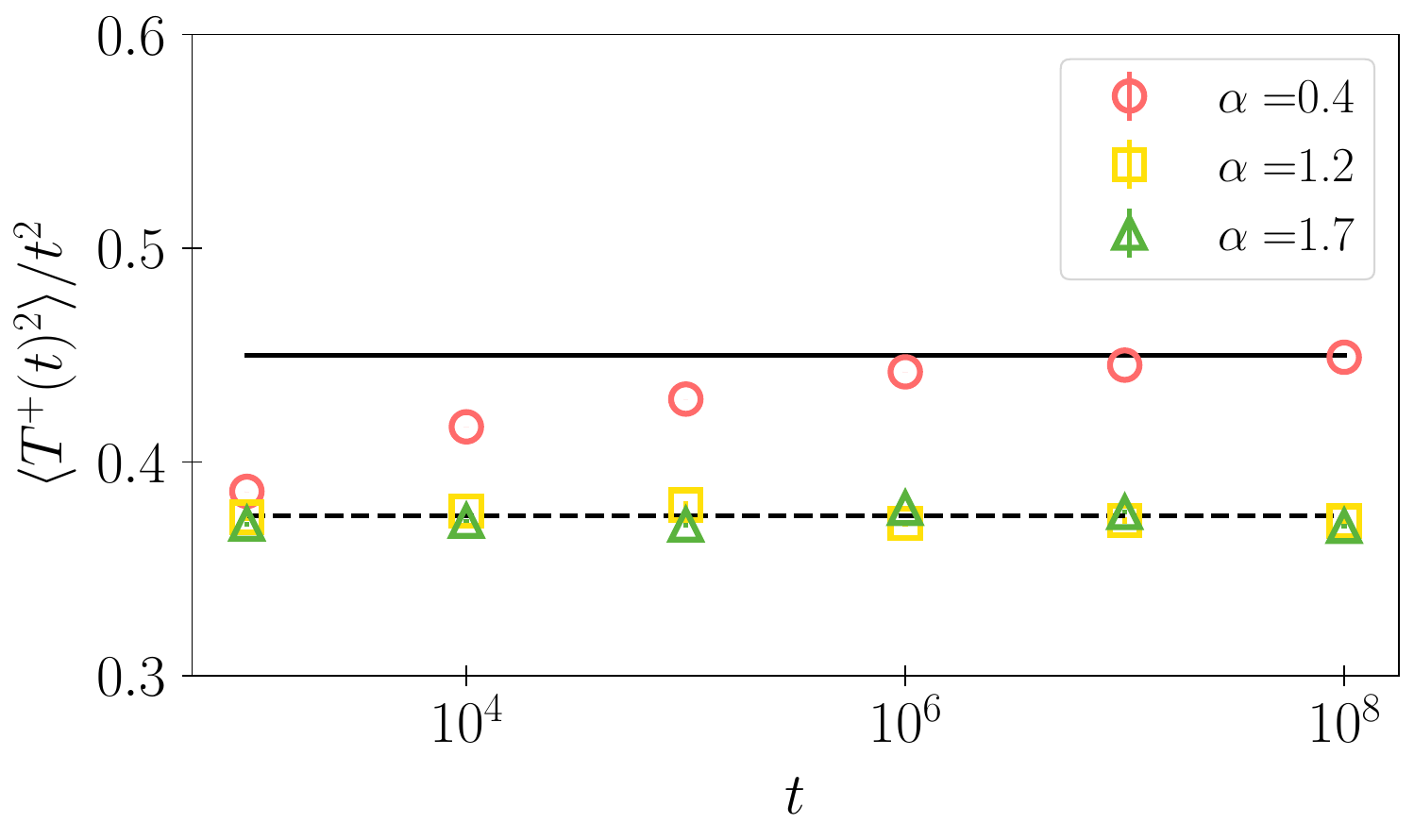}
    \caption{The second moment of the half occupation time. All the data points are computed from the numerical simulations using the waiting time PDF in Eq. \eqref{fit} (FT): the red circles for $\alpha=0.4$, the yellow squares for $\alpha=1.2$, and the green triangles for $\alpha=1.7$. The lines are theoretical and computed using Eq. \eqref{eq:T2_plus}: the solid line for $\alpha=0.4$ and the dashed line for $1<\alpha<2$. The parameters used for the numerical simulations are $x_0=0$, $\sigma=0.01$, $t_0=1$, and the number of walkers for the simulation $N=10^5$.}
    \label{fig:T2_plus}
\end{figure}

However, this universal behavior has not been observed for the second moment. Taking the second derivative of Eq. \eqref{QTP} with respect to $p$ and taking $p=0$ 
we find the general expression in Laplace space
\begin{eqnarray}
    \left\langle T^{+}(s)^{2}\right\rangle =\frac{1}{4s^{3}}\left[3+\frac{s}{K(s)}\frac{dK(s)}{ds}\right].
    \label{t2mas}
\end{eqnarray}
For a waiting time PDF with finite moments $\left\langle T^{+}(s)^{2}\right\rangle \simeq 3/4s^3$ so that
\begin{eqnarray}
    \left\langle T^{+}(t)^{2}\right\rangle _{\rm{FM}} \simeq \frac{3t^{2}}{8},
    \label{tm2ft}
\end{eqnarray}
which implies that second moment of the half occupation time grows quadratically in time. For the case of a walker performing rests of power-law distributed duration, using the expression in Eq. \eqref{Kap} for $K(s)$ and substituting in Eq.\eqref{t2mas}, one finds
\begin{eqnarray}
    \left\langle T^{+}(t)^{2}\right\rangle _{\rm{FT}}\simeq\left\{ \begin{array}{ll}
\frac{4-\alpha}{8}t^{2}, & 0<\alpha<1\\
\frac{3}{8}t^{2}, & 1<\alpha<2
\end{array}\right.
\label{eq:T2_plus}
\end{eqnarray}
in the long time limit. The same result for $0<\alpha<1$ is obtained in the case of the ML PDF. Inserting Eq. \eqref{MLlt} into Eq. \eqref{t2mas} we find
$$
\left\langle T^{+}(t)^{2}\right\rangle _{\rm{ML}}=\frac{4-\alpha}{8}t^{2}.
$$
In this case it is possible to obtain the $n$-order moment $\left\langle T^{+}(t)^{2}\right\rangle$ both for FM and ML waiting times
\begin{eqnarray}
    \left\langle T^{+}(t)^{n}\right\rangle _{\textrm{FM}}\simeq\frac{\Gamma\left(n+\frac{1}{2}\right)}{n!\sqrt{\pi}}t^{n}
\end{eqnarray}
in the long time limit and 
\begin{eqnarray}
    \left\langle T^{+}(t)^{n}\right\rangle _{\textrm{ML}}=\frac{\mathcal{B}_{n}(\alpha)}{n!}t^{n}
    \label{tnht}
\end{eqnarray}
with
$$
\mathcal{B}_{n}(\alpha)=\sum_{j=0}^{\infty}(-1)^{j}\left[\left(1+\frac{\alpha j}{2}\right)_{n}+\left(\frac{\alpha}{2}+\frac{\alpha j}{2}\right)_{n}\right]
$$
where $(a)_n=\Gamma(a+n)/\Gamma(a)$ is the Pochhammer symbol. The details of the derivation can be found in Appendix \ref{Appendix-I}. The second moment in Eq. \eqref{eq:T2_plus} is verified with numerical simulations in Fig. \ref{fig:T2_plus}. In this plot we note that $\left\langle T^{+}(t)^{2}\right\rangle$ is proportional to $t^2$ in the long time limit. As predicted in Eq. \eqref{eq:T2_plus} the coefficient is the same for $\alpha=1.2$ and $\alpha=1.7$.

\subsection{Ergodic properties}
We now study the ergodic properties of the observable $\theta (x(t))$ whose time average is $T^+/t$
$$
\overline{\theta(x(t))}:=\frac{1}{t}\int_{0}^{t}\theta(x(t'))dt'=\frac{T^{+}(t)}{t}.
$$
Then, $\left\langle \overline{\theta(x(t))}\right\rangle =\left\langle T^{+}(t)\right\rangle /t$ and $\left\langle \overline{\theta(x(t))}^{2}\right\rangle =\left\langle T^{+}(t)^{2}\right\rangle /t^{2}$. So that, from Eq. \eqref{EB}
\begin{eqnarray}
    \textrm{EB}=\lim_{t\to\infty}\frac{\left\langle T^{+}(t)^{2}\right\rangle -\left\langle T^{+}(t)\right\rangle ^{2}}{\left\langle T^{+}(t)\right\rangle ^{2}}.
    \label{EB+}
\end{eqnarray}
Inserting Eqs. \eqref{eq:halfoctime} and \eqref{tm2ft} for FM together with Eqs. \eqref{eq:halfoctime} and \eqref{eq:T2_plus} for FT into Eq. \eqref{EB+} we find
$$
\textrm{EB}=\left\{ \begin{array}{cc}
1-\frac{\alpha}{2}, & 0<\alpha<1\\
\frac{1}{2}, & 1<\alpha<2.
\end{array}\right.
$$
In consequence, the half occupation time for both Markovian and non-Markovian random walk also turns out to be non ergodic.

%\textcolor{red}{AP: In the earlier version, there was one example of run and tumble walker. In the CTRW framework
%the telegrapher's equation can be obtained by considering the memory
%kernel $K(t)=e^{-t/\tau}/\tau^{2},$ so that $K(s)=1/[\tau(1+s\tau)].$ The waiting time for the RTP turns out to be $\varphi(t)=\frac{2 e^{-\frac{t}{2 \tau }} \sin
%   \left(\frac{\sqrt{3} t}{2 \tau
%   }\right)}{\sqrt{3} \tau }$. Can this be derived from the dynamics of RTP or is it derived from the telegraphic process?}
%\textcolor{blue}{VM: We do not include the RT process here because it can not be described by a continuous-time random walk. Note that the waiting time PDF can be negative and this has no physical meaning} \textcolor{red}{AP: Yes, I saw that. Thanks. We will take out this part. }

\section{Occupation time under stochastic resetting}
\label{OT-resetting}
In this section we consider the occupation time statistics of the random walker which is further subjected to a stochastic resetting mechanism. Generically, resetting dynamics reverts the position of the walker to the starting point $x_0$. It has been shown that such intermittent dynamics can showcase a plethora of non-equilibrium and first passage optimization phenomena. We refer to \cite{EvMaSh20,pal2023random,gupta2022stochastic} for an extensive review of this topic. 

If the resetting is a Poisson process i.e., the inter-waiting times between resetting are drawn from an exponential distribution with mean $1/r$ (or equivalently, resetting occurs at a rate $r$) then the generating function of the occupation time $Q_r(p,s)$ can be expressed in terms of the generating function $Q(p,s)$ of the underlying reset-free process through the following renewal relation in the Laplace space \cite{Ho19} (in fact such a relation can be derived generically for a stochastic path functional along the resetting trajectory -- see \cite{MeSaTo15,pal2019local})
\begin{eqnarray}
Q_{r}(p,s)=\frac{Q(p,s+r)}{1-rQ(p,s+r)},
\label{Qr}
\end{eqnarray}
where we have assumed that the starting point is the origin. This is an important relation since one can compute the moments of the resetting process from the knowledge of $Q(p,s)$ and by making use of Eq. \eqref{Qr}.

\subsection{Moments -- General expressions and asymptotics}
We start by redefining the stochastic path functional along a random trajectory under resetting in the following way
\begin{eqnarray}
    Z(t)_r=\int _0 ^t U[x(t')]_rdt',
    \label{Zr}
\end{eqnarray}
where $U[x]_r=\theta (-a<x<a)$ or $U[x]_r=\theta (x)$ stands for the occupation time in an interval and the half occupation time, respectively. 
%We can find the two first moments in terms of the moments when the underlying random walk is not affected by resetting. 
The mean value of the functional \eqref{Z} i.e., the first moment of the PDF $Q_r(Z,t)$ can be written in Laplace space, with $x_0=0$, following Eq. \eqref{moments}
\begin{eqnarray}
    \left\langle Z(s)\right\rangle _{r}=-\left(\frac{\partial Q_{r}(p,s)}{\partial p}\right)_{p=0}.
    \label{Z1}
\end{eqnarray}
We can now use the renewal equation
\eqref{Qr} and the relation \eqref{Z1} to represent the mean in the following way
\begin{align}
\left\langle Z(s)\right\rangle _{r}=\frac{\left\langle Z(s+r)\right\rangle }{\left[1-rQ(p=0,s+r)\right]^{2}}.    
\end{align}
From Eq. \eqref{Qps} and setting $p=0$ one has $Q(p=0,s)=s^{-1}$, so that $Q(p=0,s+r)=(s+r)^{-1}$ -- substituting which in the above, one finds
\begin{eqnarray}
    \left\langle Z(s)\right\rangle _{r}=\left(\frac{s+r}{s}\right)^{2}\left\langle Z(s+r)\right\rangle ,
    \label{z1s}
\end{eqnarray}
which connects the first moment between the two processes. One can try to extract the long time limit by setting 
$s\to 0$ which implies $\left\langle Z(s)\right\rangle _{r}\simeq r^{2}\left\langle Z(s=r)\right\rangle /s^{2}$ and thus, by Laplace inversion with respect to $s$, one can write
\begin{eqnarray}
    \left\langle Z(t)\right\rangle _{r}\simeq r^{2}\left\langle Z(s=r)\right\rangle t+O(0)\quad\textrm{as}\;t\to\infty.
    \label{z1a}
\end{eqnarray}
Proceeding similarly for the second moment we find the following general relation
\begin{eqnarray}
    \left\langle Z^{2}(s)\right\rangle _{r}&=&\left(\frac{s+r}{s}\right)^{2}\left[\left\langle Z^{2}(s+r)\right\rangle \right.\nonumber\\
    &+&\left.\frac{2r}{s}(s+r)\left\langle Z(s+r)\right\rangle ^{2}\right]
    \label{z2s}
\end{eqnarray}
which in the long time limit i.e., $s\to 0$ yields $\left\langle Z^{2}(s)\right\rangle _{r}\simeq2r^{4}\left\langle Z(s=r)\right\rangle ^{2}s^{-3}$,  so that
\begin{eqnarray}
    \left\langle Z^{2}(t)\right\rangle _{r}\simeq r^{4}\left\langle Z(s=r)\right\rangle ^{2}t^{2}+O(t) , \quad\textrm{as}\;t\to\infty
    \label{z2a}
\end{eqnarray}
after Laplace inversion. Now we compute the variance of $Z(t)_r$ in the presence of resetting. From Eqs. \eqref{z1a} and \eqref{z2a} we see that the terms $O(t^2)$ cancel and the variance grows linearly with time in the large time limit. In particular, from Eqs.  \eqref{z1s} and \eqref{z2s} and expanding in Taylor series for small $s$ we readily find
\begin{eqnarray}
   \textrm{Var}(Z(t))_r&=& \left\langle Z^{2}(t)\right\rangle _{r}-\left\langle Z(t)\right\rangle _{r}^{2}\nonumber\\
   &=&r^{2}At+O(0),\quad\textrm{as}\;t\to\infty
   \label{Var}
\end{eqnarray}
where
\begin{eqnarray}
  A&=&\left\langle Z^{2}(s=r)\right\rangle +2r\left\langle Z(s=r)\right\rangle ^{2}\nonumber\\
  &+&2r^{2}\left\langle Z(s=r)\right\rangle \left(\frac{d\left\langle Z(s)\right\rangle }{ds}\right)_{s=r}.  
  \label{A}
\end{eqnarray}
%Note that the form in which the moments depend explicitly on time in the long time limit does not depend on the underlying random walk.
It should be noted that the explicit time dependence of the moments do not really depend on the details of the underlying random walk, atleast in the long time limit. 

\subsubsection{Occupation time}
Let us now consider the case of the occupation time in the full interval. Using Eq.
\eqref{m1}, Eqs.  \eqref{z1a} and \eqref{z2a} read
\begin{eqnarray}
    \left\langle T_{a}(t)\right\rangle _{r}\simeq\left(1-e^{-a\sqrt{\frac{2r}{\sigma^{2}K(r)}}}\right)t,\quad\textrm{as}\;t\to\infty
    \label{fm}
\end{eqnarray}
and
\begin{eqnarray}
    \left\langle T_{a}^{2}(t)\right\rangle _{r}\simeq\left(1-e^{-a\sqrt{\frac{2r}{\sigma^{2}K(r)}}}\right)^{2}t^{2},\quad\textrm{as}\;t\to\infty
    \label{sm}
\end{eqnarray}
respectively. The first and second moments above can be specified for the various waiting time PDFs: FM, FT and ML by inserting the corresponding expression for the memory kernel in the Laplace space evaluated at $s=r$.  For the exponential waiting time (with mean $\tau$) $K(r)=1/\tau$, for the ML waiting time $K(r)=(\tau^{*})^{-\alpha}r^{1-\alpha}$ and for the FT waiting time $K(r)$ is given by Eq. \eqref{eq:mk} with $\varphi (r)$ given by Eq. \eqref{ltft}.

The variance of the occupation time in an interval follows from Eqs. \eqref{Var} and \eqref{A} and can be cast in the form
\begin{eqnarray}
     \textrm{Var}(T_{a}(t))_{r}\simeq r^2A_at,
     \label{Var2}
\end{eqnarray}
with
\begin{eqnarray}
 A_{a}&=&\frac{2e^{-\lambda(r)}}{r^{3}}\left\{ 1-\left(1+\lambda(r)\Phi(r)\right)e^{-\lambda(r)}\right. \nonumber\\
 &-&\left.\frac{\Phi(r)}{2}\left(1-e^{-2\lambda(r)}\right)\right\} ,
 \label{Aa}
\end{eqnarray}
where $\lambda(r)$ and $\Phi(r)$ are given by Eqs. \eqref{lambda} and \eqref{Fi} and by the explicit expressions for $K(r)$ as explained below Eq. \eqref{sm}.

\subsubsection{Half occupation time}
\begin{figure}[h!]
    \centering
    \includegraphics[width=0.85\linewidth]{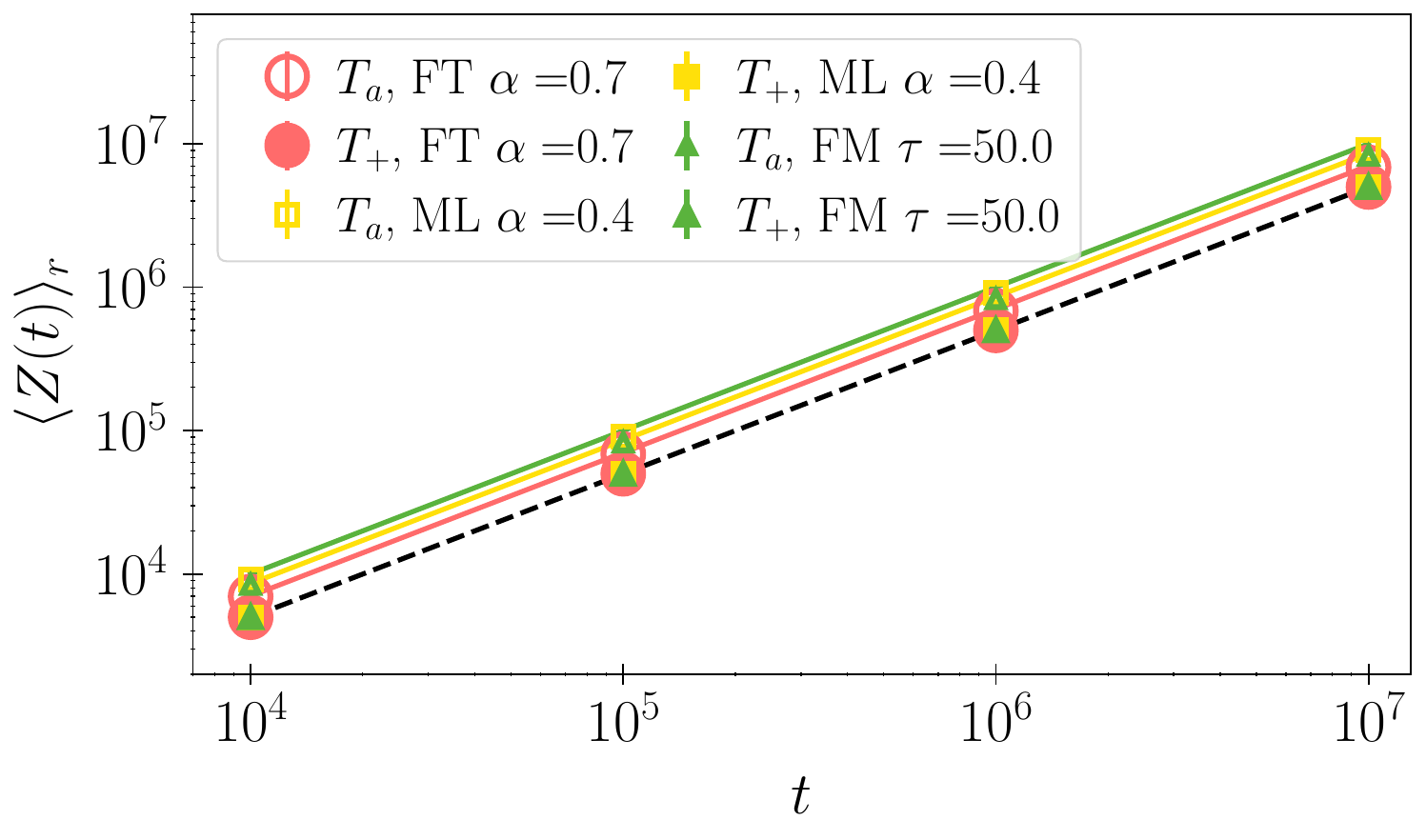}
    \includegraphics[width=0.85\linewidth]{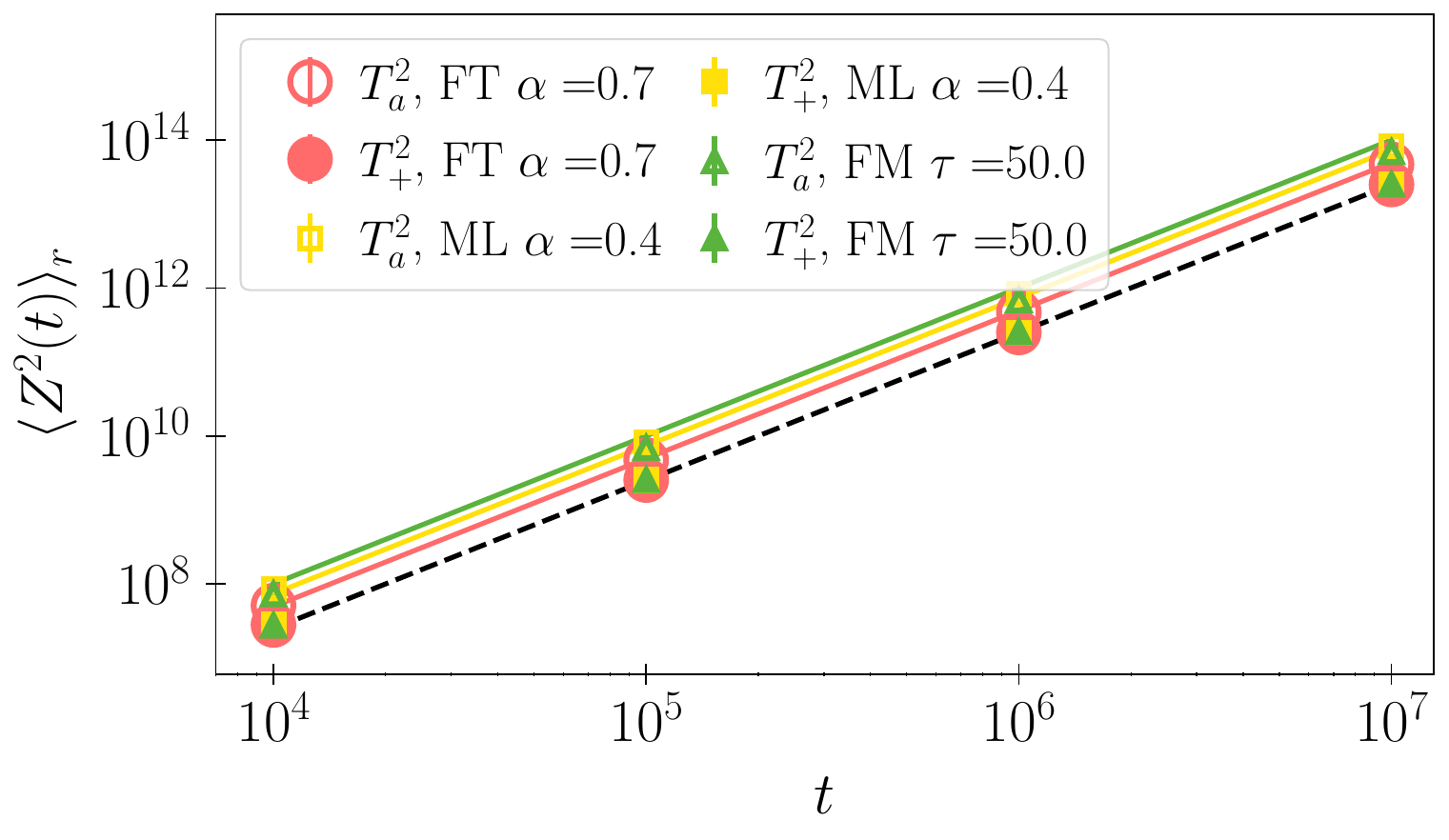}
    \includegraphics[width=0.85\linewidth]{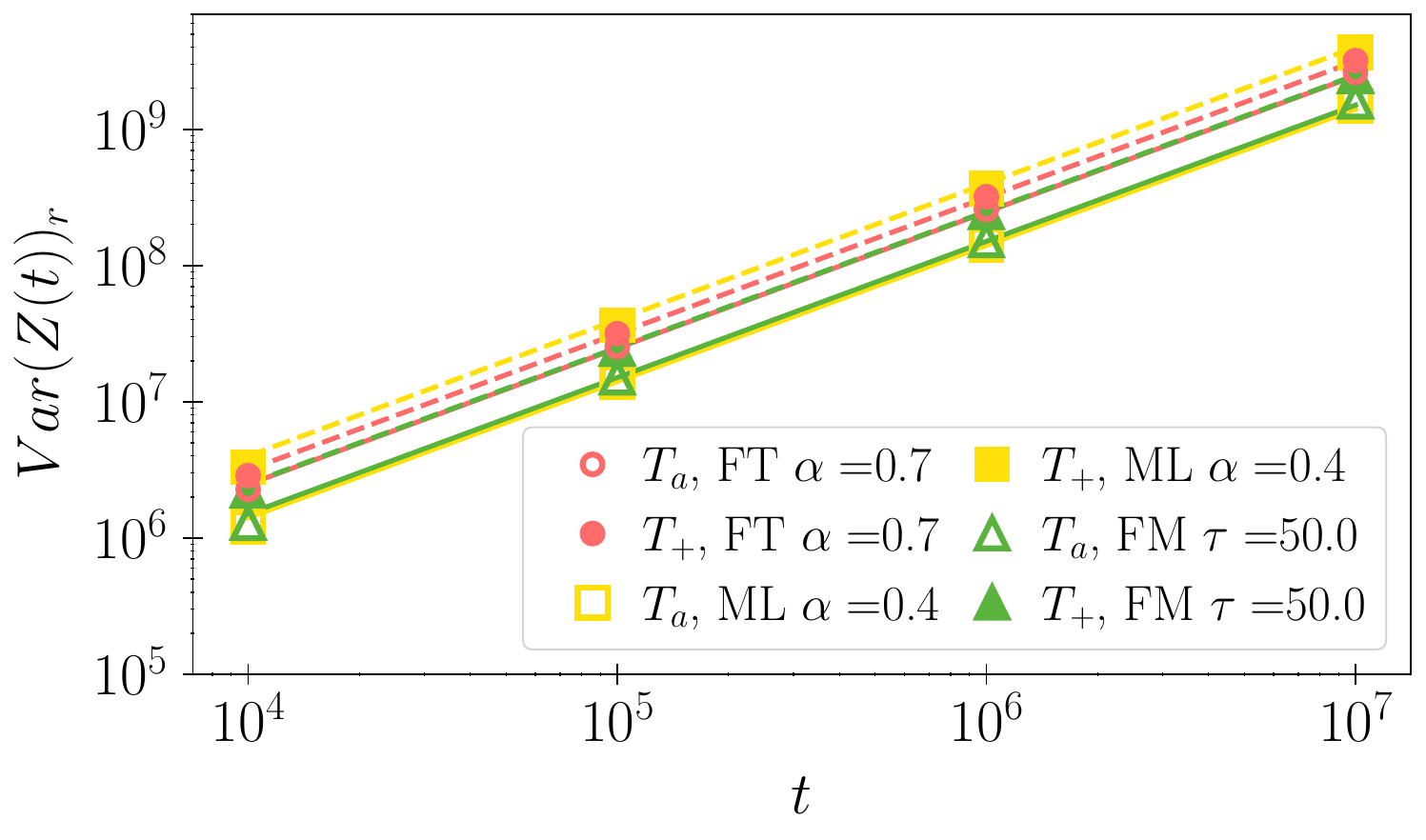}
    \caption{First moment (top panel), second moment (middle panel), and variance (bottom panel) of the occupation time in an interval $T_a$ (empty symbols), and of the half occupation time, $T^+$ (full symbols) in the presence of stochastic resetting, with reset rate $r=0.001$. The red circles (FT) are computed with the waiting time PDF in Eq. \eqref{fit} with $\alpha=0.7$ and $t_0=1$, the yellow squares (ML) with the waiting time PDF in Eq. \eqref{FII} with $\alpha=0.4$ and $\tau^*=|\Gamma(1-\alpha)|^{1/\alpha}$, and the green triangles (FM) with an exponential PDF $\varphi(t)= e^{-t/\langle \tau \rangle }/\langle \tau \rangle $ with $\langle \tau \rangle=50$. Theoretical predictions for the occupation time: The solid lines represent Eq. \eqref{fm} for the occupation time interval in the top panel, Eq. \eqref{sm} in the middle panel, and Eq. \eqref{Var2}  in the bottom panel; the colors correspond to the specific waiting time of the matching colored data points. The dashed lines are theoretical and represent the half occupation time: namely, Eq. \eqref{m3} in the top and middle panels and Eq. \eqref{Var3} in the bottom panel. The parameters used for the numerical simulations are $a=0.055$, $x_0=0$, $\sigma=0.01$, and the number of walkers for the simulation $N=10^6$. All plots are in log-log scale.}
    \label{fig:Z_r}
\end{figure}

Now we state the results given in \eqref{z1a}, \eqref{z2a} and \eqref{Var} for the half occupation time. Considering $\left\langle T^{+}(s)\right\rangle =1/2s^{2}$ and using Eq. \eqref{t2mas} we find the following expression for the first two moments
\begin{eqnarray}
\left\langle T^{+}(t)\right\rangle _{r}\simeq\frac{t}{2},\quad\left\langle T^{+}(t)^{2}\right\rangle _{r}\simeq\frac{t^{2}}{4},
    \label{m3}
\end{eqnarray}
and the variance follows
\begin{eqnarray}
\textrm{Var}(T^{+}(t))_{r}\simeq r^2A_+t,
\label{Var3}
\end{eqnarray}
with
\begin{eqnarray}
   A_+= \frac{1}{4r^3}\left[1+\frac{r}{K(r)}\frac{dK(r)}{dr}\right].
   \label{Am}
\end{eqnarray}
In Figure \ref{fig:Z_r} we compare the expressions for the first moment, the second moment, and the variance of both the occupation time of an interval and the half occupation time against numerical simulations. Since the first and second moments scale as a power of time they appear as linear functions in the log-log scale of Fig. \ref{fig:Z_r}. Moreover, the slopes of the lines in each panel are the same, which indicates that the temporal scaling is independent of the underlying random walk.

\subsection{PDF}
\begin{figure}[h!]
    \centering
    \includegraphics[width=\linewidth]{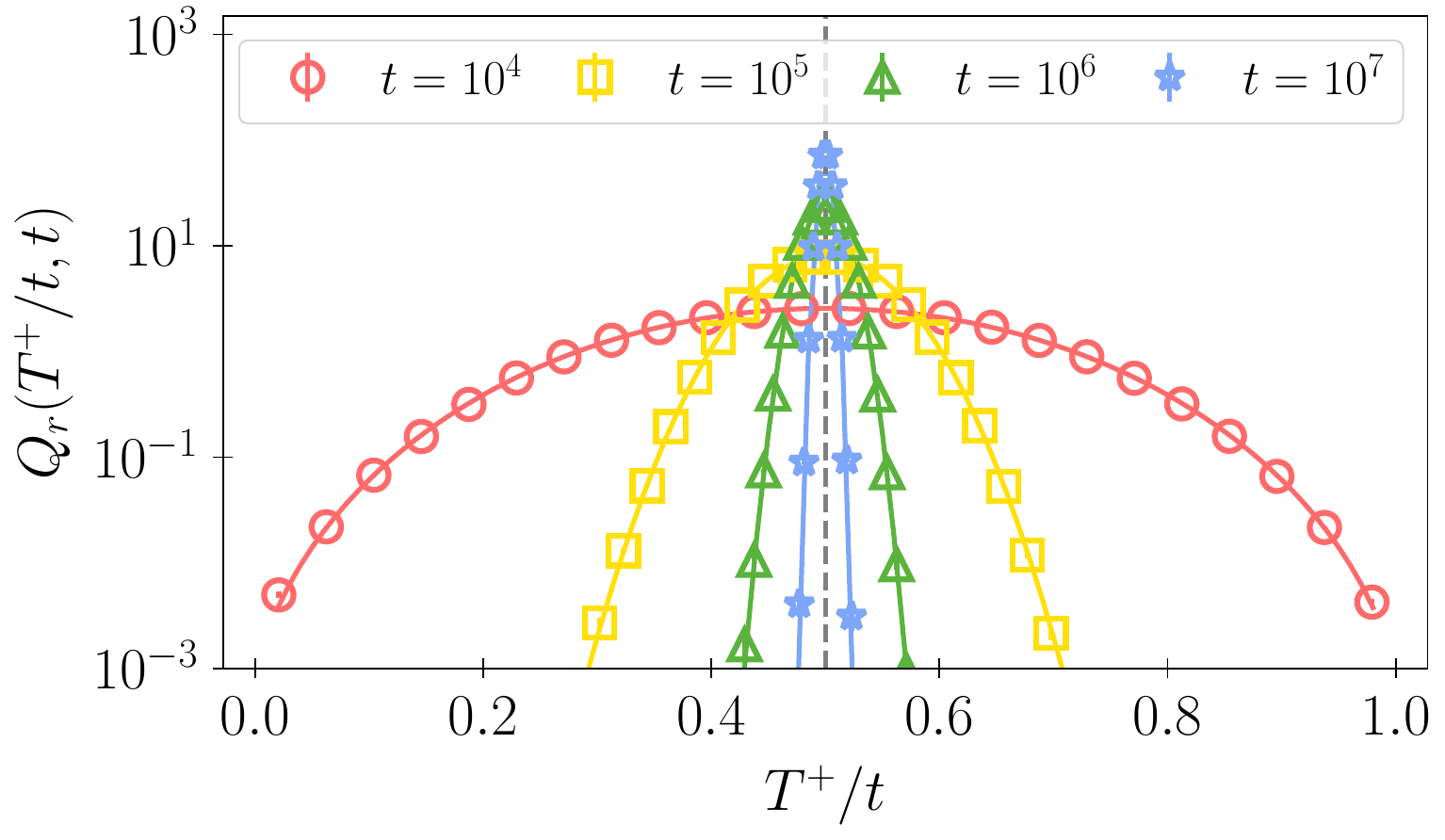}
    \caption{Scaled PDF $Q_r(T^+/t,t)$ of the half occupation time in the presence of stochastic resetting for an exponential waiting time PDF. The data is computed for four evolution times: $t=10^4$ red circles, $t=10^5$ yellow squares, $t=10^6$ green triangles, and $t=10^7$ blue stars. The solid lines are theoretical and computed using Eq. \eqref{Eq:exp_r}. The vertical dashed line represents $\langle T^+ (t) \rangle_r/t$, from Eq. \eqref{m3} around which all the PDFs are centered. The parameters used for the numerical simulations are $\tau=50$, $x_0=0$, $\sigma=0.01$, $r=0.001$ and the number of walkers for each simulation $N=10^6$. To compute $T^+$, we add $1/2$ of the residence time at $x=0$ to the total time in the positive axis. The plot is in log-linear scale.}
    \label{fig:Q_r_tpl_exp}
\end{figure}

In this section, we present the results for the PDF of the occupation times under resetting dynamics. To start with, we first compute the generating function of the PDF $Q_r(p,s)$ for the occupation time in the presence of resetting. This is related to the generating function of the PDF without resetting $Q(p,s)$ through the relation  \eqref{Qr} in the Laplace space. First we note that $Q(p,s)$ in Eq. \eqref{Qr} is shifted in the Laplace space so that
\begin{eqnarray}
Q_{r}(T,t)=e^{-rt}\mathcal{L}_{p\to T}^{-1}\mathcal{L}_{s\to t}^{-1}\left[\frac{Q(p,s)}{1-rQ(p,s)}\right],
    \label{Qrr}
\end{eqnarray}
where $T$ represents either the occupation time in an interval $T_a$
 or the half occupation time $T^+$; $t$ represents the observation time and the conjugate parameters are self-explanatory. Performing the double Laplace transform
is not possible in the case of the occupation time in an interval, i.e., when $Q(p,s)$ is given by Eq. \eqref{QTA}. However, for the half occupation time we can proceed further. First we note that the variable $p$ in Eq. \eqref{QTP} is shifted by $s$. Then, applying the shifting property in the Laplace space 
\begin{align}
    \mathcal{L}_{p\to T^{+}}^{-1}\left[\frac{Q(p,s)}{1-rQ(p,s)}\right]=e^{-sT^{+}}\mathcal{L}_{p\to T^{+}}^{-1}\left[\frac{\psi(p,s)}{1-r\psi(p,s)}\right],
        \label{psi}
\end{align}
where $\psi(p,s)\equiv Q(p-s,s)$. Using the  series expansion
\begin{eqnarray}
    \frac{\psi(p,s)}{1-r\psi(p,s)}=\sum_{k=0}^{\infty}r^{k}\psi(p,s)^{k+1},
\end{eqnarray}
together with Eq. \eqref{psi} followed by some rearrangement, Eq. \eqref{Qrr} yields
\begin{align}
Q_{r}(T^{+},t)=e^{-rt}\sum_{k=0}^{\infty}r^{k}\mathcal{L}_{p\to T^{+}}^{-1}\mathcal{L}_{s\to t-T^{+}}^{-1}\left[\psi(p,s)^{k+1}\right].
\end{align}
To proceed further we need to specify the expression for the memory kernel. If we consider the ML waiting time PDF \eqref{MLlt} then
$$
\psi(p,s)=\frac{1+s^{1-\frac{\alpha}{2}}p^{\frac{\alpha}{2}-1}}{s+s^{1-\frac{\alpha}{2}}p^{\frac{\alpha}{2}}}=\frac{1}{s}\frac{1+\left(\frac{p}{s}\right)^{\frac{\alpha}{2}-1}}{1+\left(\frac{p}{s}\right)^{\frac{\alpha}{2}}}.
$$
To invert with respect to $p$ we make use of the scaling property of the Laplace transform to get
\begin{align}
    Q_{r}(T^{+},t)&=e^{-rt}\sum_{k=0}^{\infty}r^{k}\mathcal{L}_{s\to t-T^{+}}^{-1}
    %\left\{ \right.
    \nonumber\\
    & \left\{\frac{1}{s^{k+1}}\mathcal{L}_{p\to sT^{+}}^{-1}\left[\left(\frac{1+p^{\frac{\alpha}{2}-1}}{1+p^{\frac{\alpha}{2}}}\right)^{k+1}\right]\right\} ,
\end{align}
where the details of the derivation has been moved to the Appendix \ref{Appendix-I}. Note that when $\alpha=1$ we have to recover the case when the waiting time is exponential. On setting $\alpha=1$ one has $\psi(p,s)=p^{-1/2}$ and the double Laplace inversion can be easily found. We finally find in this case
\begin{align}
    Q_{r}(y,\xi)&=\xi e^{-\xi}\sum_{k=0}^{\infty}\frac{\left[\xi^{2}y(1-y)\right]^{\frac{k-1}{2}}}{\Gamma\left(\frac{k+1}{2}\right)^{2}}=\frac{e^{-\xi}}{\pi\sqrt{y(1-y)}}\nonumber\\
    &+\xi e^{-\xi}\left[L_{0}\left(2\xi\sqrt{y(1-y)}\right)+I_{0}(2\xi\sqrt{y(1-y)})\right],
    \label{Eq:exp_r}
\end{align}
where $L_0(\cdot)$ and $I_0(\cdot)$ are the modified Struve and Bessel functions respectively and $\xi=rt$. This result holds for all $t$ and agrees with the result found in Ref. \cite{Ho19}. We confirm it with numerical simulations in Fig. \ref{fig:Q_r_tpl_exp}. 

The exact solution for FT or ML waiting time turns out to be intractable. 
 However we can find the limiting distribution in the long time limit. To this end we express $ Q_r(p,s)$ in a moment generating form 
\begin{eqnarray}
    Q_{r}(p,s)=\sum_{n=0}^{\infty}\frac{(-p)^{n}}{n!}\left\langle T_{a}^{n}(s)\right\rangle _{r},
    \label{mgf}
\end{eqnarray}
where the expression for the moments $\left\langle T_{a}^{n}(s)\right\rangle _{r}$ in the long time limit can be easily found. From Eq. \eqref{z1s} we have $\left\langle T_a(s)\right\rangle _{r}\simeq r^{2}\left\langle T_a(r)\right\rangle /s^{2}$ as $s\to 0$. From Eq. \eqref{z2s} the second moment follows analogously, $\left\langle T_{a}^{2}(s)\right\rangle _{r}\simeq2r^{4}\left\langle T_{a}(r)\right\rangle ^{2}s^{-3}$. In general, it can be shown from Eq. \eqref{QTA} that $\left\langle T_{a}^{n}(s)\right\rangle _{r}\simeq n!r^{2n}\left\langle T_{a}(r)\right\rangle ^{n}s^{-n-1}$. Thus, from Eq. \eqref{mgf}, it follows
$$
Q_{r}(p,s)\simeq\frac{1}{s}\sum_{n=0}^{\infty}\left(-\frac{pr^{2}}{s}\left\langle T_{a}(r)\right\rangle \right)^{n},
$$
and inverting $s\to t$, we find
$$
Q_{r}(p,t)\simeq\sum_{n=0}^{\infty}\frac{1}{n!}\left(-pr^{2}\left\langle T_{a}(r)\right\rangle t\right)^{n}=e^{-pr^{2}\left\langle T_{a}(r)\right\rangle t}.
$$
Finally, inverting $p\to T_a$
\begin{eqnarray}
    Q_{r}(T_a,t)\simeq\delta\left(T_{a}-\left\langle T_{a}(t)\right\rangle _{r}\right),
    \label{ltl}
\end{eqnarray}
where $\left\langle T_{a}(t)\right\rangle _{r}$ is given by Eq. \eqref{fm}. This means that in the long time limit (i.e., when there have been a sufficiently large number of resetting events $rt \gg 1$) the occupation time in the interval $[-a,a]$ is given by Eq. \eqref{fm}. This is expected because the occupation fraction or the empirical density namely $ \lim _{t \to \infty} T_{a}/t$, will converge to the steady state probability of the particle to be in the interval. This probability is $\int_{-a}^{a}P(x,t)_{r}dx$ where $P(x,t)_{r}$ is the walker's propagator in the presence of resetting. From Ref. \cite{MaCaMe19} this propagator reaches a steady state and is given by
$$
P_r^{ss}(x)=P_{r}(x,t\to\infty)=\sqrt{\frac{r}{2\sigma^{2}K(r)}}e^{-|x|\sqrt{\frac{2r}{\sigma^{2}K(r)}}}.
$$
Hence,
$$
 \lim _{t \to \infty} \frac{ T_{a}}{t}=\int_{-a}^{a}P_{r}(x)dx=1-e^{-a\sqrt{\frac{2r}{\sigma^{2}K(r)}}}.
$$
Proceeding analogously for the half occupation time, and computing the higher order moments as to find \eqref{m3} we infer that $\langle T^{+}(t)^{n}\rangle_{r}\simeq(t/2)^{n}$ and 
$$
Q_{r}(p,t)=\sum_{n=0}^{\infty}\frac{(-p)^{n}}{n!}\left\langle T^{+}(t)^{n}\right\rangle _{r}\simeq e^{-pt/2},
$$
and Laplace inverting the PDF of the half occupation time under resetting is given by 
\begin{eqnarray}
    Q_{r}(T^+,t)\simeq \delta (T^+-t/2). 
    \label{ltl2}
\end{eqnarray}
This result is expected because the half occupation fraction  $\lim _{t \to \infty} T^+/t$, will converge to the steady state probability of the particle to be in the half line $x>0$. This probability is given by $\int_{0}^{\infty}P^{ss}_{r}(x)dx=1/2$. Hence, we have
\begin{align}
    \lim _{t \to \infty} \frac{T^+}{t}=\int_{0}^{\infty}P_{r}^{ss}(x)dx=\frac{1}{2}.
\end{align}

We note in passing, as suggested by 
Eqs. \eqref{ltl} and \eqref{ltl2}, that the PDF of any stochastic functional (without having explicit time dependence on an external parameter) in the presence of Poissonian resetting should attain the form $Q_r(Z,t)\simeq \delta (Z-   \left\langle Z(t)\right\rangle _{r})$ in the long time limit although a rigorous proof of this claim is beyond the scope of this article. 
%In Figure \ref{fig:Q_r_t} we present numerical simulations of  $Q(Z/t,t)$ for both the occupation time on an interval, top panel, and the half occupation time, bottom panel. In both cases, we can see how $Q(y,t)$ converges to $ \delta(y - \langle Z(t) \rangle_r/t)$ in the long time limit, where $y=Z/t$. 

%\begin{figure}[h!]
%    \centering
%    \includegraphics[width=\linewidth]{Q_r_Tat-alpha_0.4}
%    \includegraphics[width=\linewidth]{Q_r_Tplt-alpha_0.4}
%    \caption{PDF $Q(Z/t,t|0)_{r}$ of the occupation time on an interval $Z=T_a$ (top panel), and the half occupation time $Z=T^+$ (bottom panel) in the presence of stochastic resetting for the FT waiting time PDF, Eq. \eqref{fit}. The data is computed for four evolution times: $t=10^4$ red circles, $t=10^5$ yellow squares, $t=10^6$ green triangles, and $t=10^7$ blue stars. The vertical dashed line represents $\langle Z (t) \rangle_r$, Eq. \eqref{fm} top panel, and \eqref{m3} bottom panel. The dotted lines are included in the top panel as a visual guide. The solid lines are computed in the bottom panel with Eq. \eqref{Ga}. The parameters used for the numerical simulations are $\alpha=0.4$, $a=0.055$, $x_0=0$, $\sigma=0.01$, $r=0.001$ and the number of walkers for each simulation $N=10^6$. To compute $T^+$, we add $1/2$ of the residence time at $x=0$ to the total time in the positive axis.  Both panels are plotted in a log-lin scale.}
 %   \label{fig:Q_r}
%\end{figure}

\begin{figure}[h!]
    \centering
    \includegraphics[width=\linewidth]{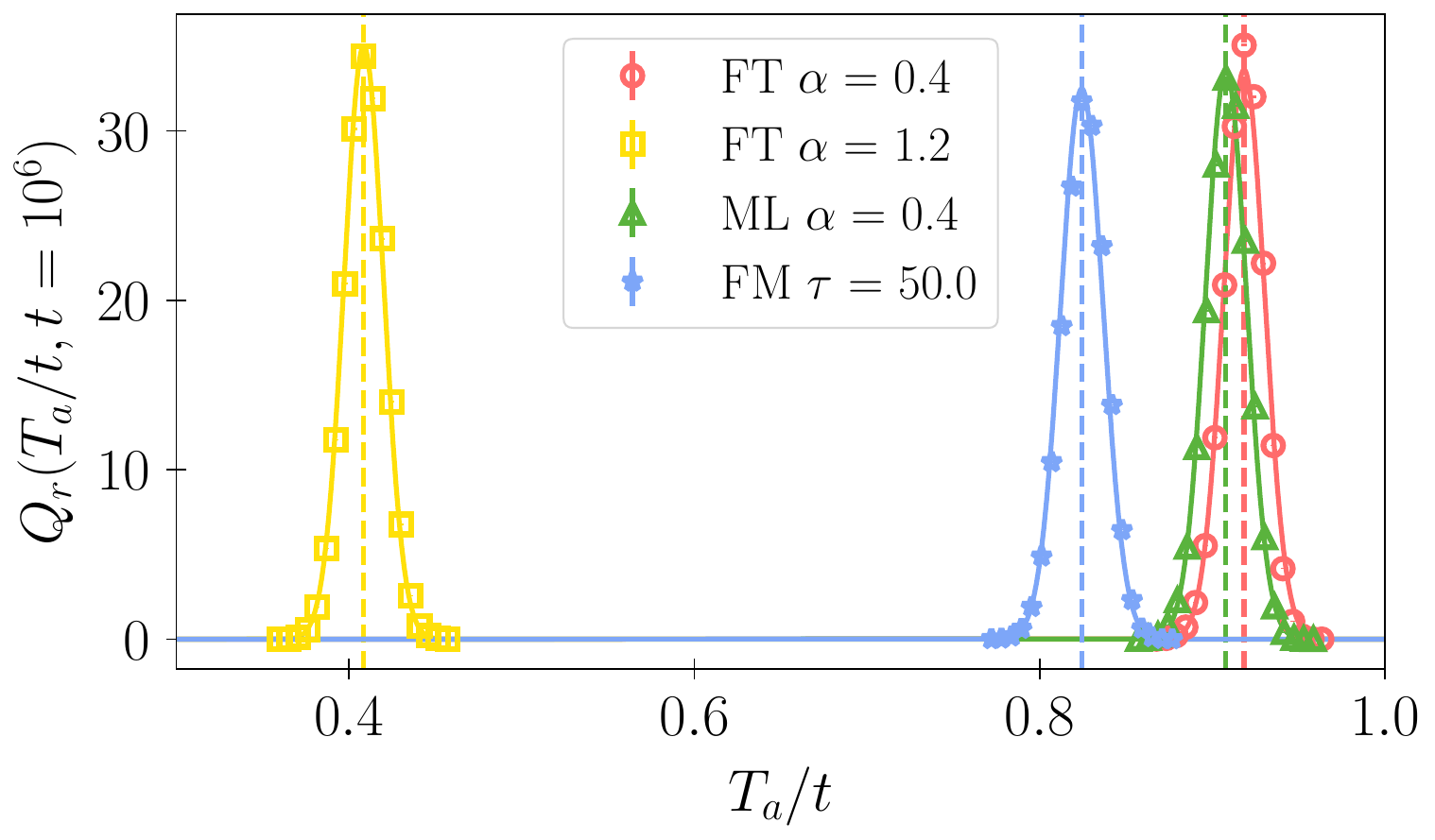}
    \includegraphics[width=\linewidth]{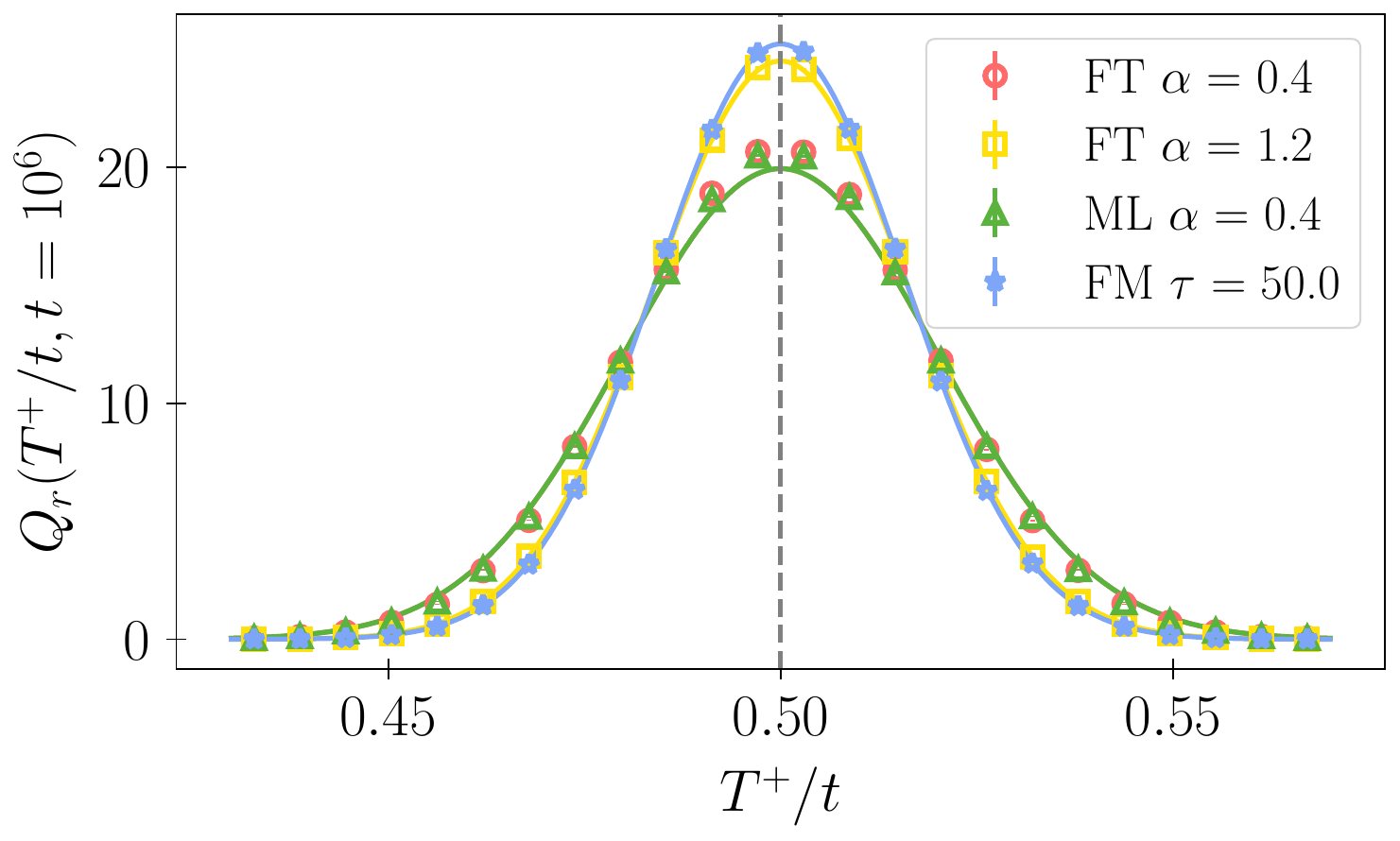}
    \caption{Scaled PDF $Q_r(Z/t,t)$ of the occupation time on an interval $Z=T_a$ (top panel), and the half occupation time $Z=T^+$ (bottom panel) in the presence of stochastic resetting. The red circles and yellow squares (FT) are from numerical simulations and are computed using the waiting time PDF in Eq. \eqref{fit} with $t_0=1$ and $\alpha=0.4$ and $\alpha=1.2$ respectively; the green triangles (ML) with the waiting time PDF in Eq. \eqref{FII} with $\alpha=0.4$ and $\tau^*=|\Gamma(1-\alpha)|^{1/\alpha}$; and the blue stars (FM) with an exponential PDF $\varphi(t)= e^{-t/\langle \tau \rangle}/\langle \tau \rangle $ with $\langle \tau \rangle=50$. The vertical dashed lines represent the means $\langle Z (t) \rangle_r/t$, Eq. \eqref{fm} top panel, and Eq. \eqref{m3} bottom panel. Theoretical predictions: The solid lines are plotted using Eq. \eqref{Qar} in the top panel and Eq. \eqref{Qmr} in the bottom panel with their respective mean and variance. The parameters used for the numerical simulations are $a=0.055$, $x_0=0$, $\sigma=0.01$, $r=0.001$, $t=10^6$ and the number of walkers for each simulation $N=10^6$. To compute $T^+$, we added $1/2$ of the residence time at $x=0$ to the total time in the positive axis.}
    \label{fig:Q_r_t}
\end{figure}

\subsection{Ergodic properties}
This section discusses the ergodic properties of the occupation times under resetting dynamics. 
Let $x(\tau)_r$ be the trajectory of the walker starting at $\tau=0$ up to time $t$ under resetting and consider an observable $\mathcal{O}[x(t)]_r=U[x(t)]_r$ along with the stochastic functional \eqref{Zr} along this trajectory. Then, the time average $\overline{\mathcal{O}[x(t)]_{r}}=Z(t)_{r}/t$ is equal to the average over the trajectories $\left\langle \mathcal{O}[x(t)]_{r}\right\rangle $ when the observable is ergodic. In this case, the time average of the observable is no longer stochastic in the long time limit. As we have shown above, the limiting PDF of the time-averaged observable is a Dirac delta function centered at the mean value so that the occupation time is ergodic in the presence or exponentially distributed resetting times, i.e., for a Poissonian resetting process. It should also be noted that diffusion with stochastic resetting is ergodic at the mean level unlike their reset-free counterparts but resetting renders ergodicity breaking in the 
time-averaged-mean-squared-displacement \cite{stojkoski2022autocorrelation}. The occupation time also becomes non-ergodic when the time between resets are drawn from a power-law PDF \cite{BaFlMe24}.

This result is also found for the half occupation time of a Brownian walker in an arbitrary confining (stable) potential (see Section 2 in Ref. \cite{Ba06}). This fact stems from the idea that the effect of a confining potential (or a force field towards a point) is equivalent to a Poissonian resetting process to a point \cite{Evans_2013,pal2015diffusion}. However, this is not the case when the walker performs a subdiffusive random walk (see Section 3.1 in Ref. \cite{Ba06}). The difference relies in the fact that this equivalence can be established when the process is ergodic only.   
Moreover, due to the resetting-induced ergodicity the transient behaviour of the occupation time PDF follows a Gaussian PDF in the long time limit. 

Indeed, this is consequence of the fact that the occupation time obeys the central limit theorem.  This result is
not very surprising since each reset renews the process in its entirety. As a result, the occupation times accumulated in the consecutive intervals and the rest are statistically independent and identical to each other. 
Let $T_a^{\textrm{in}}$ and $T_a^{\textrm{out}}$ the occupation times spent by the walker inside and outside the interval $[-a,a]$. If $\tau^{\textrm{in}}_i$ and $\tau^{\textrm{out}}_i$ and denote the times spent by the walker inside and outside the interval between the $i-1$-th and $i$-th resets, respectively, then the time between two consecutive resettings is $\tau_i=\tau^{\textrm{in}}_i+\tau^{\textrm{out}}_i$ and $T_a^{\textrm{in}}=\sum_{i=1}^{n(t)}\tau^{\textrm{in}}_i$, $T_a^{\textrm{out}}=\sum_{i=1}^{n(t)}\tau^{\textrm{out}}_i$, where $n(t)$ is the number of resets up to time $t$. Since $\tau_i$ are exponentially distributed, the times $\tau^{\textrm{in}}_i$ and $\tau^{\textrm{out}}_i$ have finite moments so that the occupation time in an interval $T_a$ can be expressed a the sum of identically distributed independent random variables with finite mean. Therefore, by virtue of the central limit theorem the PDF for $T_a$ follows the Gaussian PDF. The same reasoning holds for the the half occupation time, its PDF is also a Gaussian. In general,
\begin{align}
Q_{r}(Z,t)\simeq\frac{1}{\sqrt{2\pi\textrm{Var}(Z)_{r}}}\exp\left[-\frac{\left(Z-\left\langle Z(t)\right\rangle _{r}\right)^{2}}{2\textrm{Var}(Z)_{r}}\right],
    \label{Ga}
\end{align}
where is either $Z=T_a$ or $Z=T^+$ where the general expressions for $\left\langle Z(t)\right\rangle _{r}$ and $ \textrm{Var}(Z) _{r}$ are given by Eqs. \eqref{z1a} and \eqref{Var}, respectively. Such Gaussian limit for the observable PDF was also observed earlier \cite{pal2019local,gupta2020work}.

Notably, the Gaussian regime is actually the \textit{transient solution} towards the Dirac-Delta limiting PDF at the large time by virtue of
\begin{eqnarray*}
\lim_{t\to\infty}Q_r(y,t)&=&\lim_{t\to\infty}\frac{1}{\sqrt{2\pi r^{2}A/t}}\exp\left[-\frac{\left(y-r^2\left\langle Z(r)\right\rangle \right)^{2}}{2r^{2}A/t}\right]\\
  &=&\delta\left(y-r^2\left\langle Z(r)\right\rangle \right)  ,
\end{eqnarray*}
where $y$ is either $y=T_a/t$ or $y=T^+/t$. We can specify Eq. \eqref{Ga} for both occupation times. For $T_a$, we find
\begin{eqnarray}
  Q_r(y,t)\simeq\frac{1}{\sqrt{2\pi r^{2}A_a/t}}\exp\left[-\frac{\left(y-1+e^{-\lambda (r)} \right)^{2}}{2r^{2}A_a/t}\right],
  \label{Qar}
\end{eqnarray}
where $y=T_a/t$ and
$A_a$ is given by Eq. \eqref{Aa}.
Similarly for $T^+$, we have
\begin{eqnarray}
Q_r(y,t)\simeq\frac{1}{\sqrt{2\pi r^{2}A_+/t}}\exp\left[-\frac{\left(y-\frac{1}{2}\right)^{2}}{2r^{2}A_+/t}\right],
\label{Qmr}
\end{eqnarray}
where $y=T^+/t$ and $A_+$ is given by Eq. \eqref{Am}.

In Fig. \ref{fig:Q_r_t} we reason these arguments using numerical simulations. In particular, we compare Eqs. \eqref{Qar} and \eqref{Qmr} with numerical simulations for $T_a$ and $T^+$ for the different waiting times at the long time limit $t=10^6$. We find that a Gaussian PDF can successfully represent the data, which confirms our theoretical predictions.

\section{Conclusions}
\label{Summary}
In summary, this work puts forward a comprehensive analysis of the occupation time statistics for a random walker which can be both Markovian or non-Markovian in nature. To this end, we derive the backward Feynman-Kac equation for a generic non-Markovian random walk functional, based on the Continuous-Time Random Walk framework. We analyzed the statistical properties for two particular functionals: the occupation time in an interval $T_a$ and the half occupation time $T^+$. In both cases we considered that the underlying random walk may be Markovian (the waiting time between jumps or the rest period is exponentially distributed) or non-Markovian (the waiting time PDF is a power law or a Mittag-Leffler function). In particular, we obtained the characteristic functions for such functionals in terms of the memory kernel and obtained exact and asymptotic expressions for the first two moments. 

For Mittag-Leffler and power-law PDF of waiting times, we showed that the mean $\left\langle T_{a}(t)\right\rangle$ converges to the same value in the long time limit if $0<\alpha<1$ while for the power-law PDF of waiting times with $1<\alpha<2$, it is seen that $\left\langle T_{a}(t)\right\rangle$  converges to the same value corresponding to the regular Brownian motion in the long time limit. However, the second moment $\left\langle T_{a}(t)^2\right\rangle$  for the power-law and Mittag-Leffler PDFs of waiting times does not converge to the same value. We analyzed the short time limit and showed that $\left\langle T_{a}(t)\right\rangle$ and $\left\langle T_{a}(t)^2\right\rangle$ do not depend on the underlying random walk. Although it was not possible to find an exact analytic solution for $Q(T_a,t)$ through the entire interval, we found exact solutions in the left ($T_a\ll t$), right edge ($T_a\to t$) regime and in the central bulk regime for the power-law and Mittag-Leffler PDFs of waiting times. For $T^+$, we find that its PDF converges to a limiting distribution in terms of the scaled variable  $T^+/t$ in the long time limit. Quite interestingly, this limiting distribution is found to be the \textit{arc-sine distribution} as long as the rest periods have finite moments. On the other hand, for both the power-law (with $0<\alpha<1$) and Mittag-Leffler PDFs of waiting times, the limiting PDF is found to be the Lamperti PDF. In addition, we found expressions for the general moments $\left\langle T_{a}(t)^n\right\rangle$ having rest periods with finite or infinite moments. The first part of the work concludes with a detailed analysis of the ergodic properties of the occupation times and their connection to the limiting probability distribution functions.

Finally, we studied the statistics of the occupation time functionals in the presence of stochastic resetting (i.e., exponentially distributed reset times). We observed that the moments of any functional behaves as $\langle Z(t)^{n}\rangle_{r}\sim t^{n}$. 
%With respect to the PDF of the functionals we computed the PDF for the half occupation time when the waiting time are exponentially distributed and distributed according to a Mittag-Leffler functions.
Quite intriguingly, we proved that the occupation times, in the presence of exponentially distributed resetting times, obey the central limit theorem and for both functionals and the PDFs eventually converge to the Dirac delta functions centered at the mean value in the long time limit. 

The study of the statistical properties of the occupation times of a (normal or subdiffusive) random walker is of potential interest in behavioral ecology since they provide useful information of how a walker occupies or covers a  territory \cite{Vi22}. Similarly, experiments with controlled \& tunable robots mimicking living entities can also measure the foraging territory \cite{paramanick2024uncovering} and shed light on their statistics that can be verified with our theory. 
%Although this work represents an advance in this idea there are still open problems. 
This work sets an open stage for many new theoretical research frontiers. For instance, it would be interesting to extend the Feynman-Kac formalism for non-Markovian random walkers under non-exponential resetting time PDF or time dependent resetting rates. An extension to higher dimension would also be a challenging avenue to explore in future. 
%In future work we hope to find the Feynman-Kac equation for random walk functional in the presence of resetting distributed according to other non-exponential PDFs.

\section*{Acknowledgements}
The authors acknowledge the financial support of the Ministerio de Ciencia e Innovaci\'on (Spanish government) under
Grant No. PID2021-122893NB-C22. AP acknowledges research support from the Department of Atomic Energy, India via the Soft-Matter Apex Projects. 

\appendix

\section{Derivation of the backward master equation for $Q(p,s|x_0)$}
\label{Appendix-A}
In this appendix, we derive the backward equation \eqref{FK} for the MGF $Q(p,s|x_0)$ describing the generic occupation time observable $T=\int_0^t U\left[ x(t') \right]dt'$. 
%(specifically, $T_a$ for the occupation time in an interval and  $T^+$ for the half occupation time). 
To start with, let us consider the walker
performing jumps of constant length $\sigma$. The walker is initially
at $x_{0}$ and after waiting a time $t$, randomly distributed
according to $\varphi(t)$, it performs a jump to either
$x_{0}+\sigma$ or $x_{0}-\sigma$ with the same probability or it stays
at $x_{0}$ during this time $t$. Hence, we can write the following evolution equation
\begin{eqnarray}
Q(T,t|x_{0})&=&\int_{0}^{t}\varphi(\tau)\left\{ \frac{1}{2}Q(T-\tau U(x_{0}),t-\tau|x_{0}+\sigma) \right .\nonumber\\
&+& \left . \frac{1}{2}Q(T-\tau U(x_{0}),t-\tau|x_{0}-\sigma)\right\} d\tau \nonumber\\
&+&\varphi^{*}(t)\delta\left[T-t U(x_{0})\right],\label{eq:bectrw}
\end{eqnarray}
where $\varphi^{*}(t)=\int_{t}^{\infty}\varphi(\tau)d\tau$ is the persistent probability that the walker does not perform any jump during the measurement
time $[0,t]$. Note that the quantum $t U(x_{0})$ is the contribution to the observable $T$ cumulated while being idle at $x_{0}$ during the time interval $[0,t]$ as indicated by the last term on the RHS. 
%The last term on the right hand side of describes a motionless particle, for which $T(t)=tU(x_0)$. 
Next, taking Laplace transform on the both sides of Eq.
(\ref{eq:bectrw}) with respect to $T$, we find
we find
\begin{eqnarray}
  Q(p,&t&|x_{0})=\int_{0}^{t}\varphi(\tau)e^{-p\tau U(x_{0})}\left\{ \frac{1}{2}Q(p,t-\tau|x_{0}+\sigma) \right .\nonumber\\
  &+& \left .\frac{1}{2}Q(p,t-\tau|x_{0}-\sigma)\right\} d\tau 
  +\varphi^{*}(t)e^{-ptU(x_{0})}.
  \label{eq:bectrw2}  
\end{eqnarray}
Taking another Laplace transform in Eq. \ref{eq:bectrw2} with respect to $t$, we find
\begin{eqnarray}
 Q(p,&s&|x_{0})=\varphi(s+pU(x_{0}))\left[\frac{1}{2}Q(p,s|x_{0}+\sigma) \right.\nonumber\\
 &+&\left.\frac{1}{2}Q(p,s|x_{0}-\sigma)\right]
 +\frac{1-\varphi(s+pU(x_{0}))}{s+pU(x_{0})}.
 \label{eq:bectrw3}   
\end{eqnarray}
Let us assume that the jump size $\sigma$ is smaller than any other
length scale of the system, i.e., $x_{0}\gg\sigma$, so that we expand
$Q(p,s|x_{0}\pm\sigma)$ up to $O(\sigma^{2})$ 
%(diffusive limit)
to arrive at
\begin{eqnarray}
 Q(p,s|x_{0}\pm\sigma)&\simeq& Q(p,s|x_{0})\pm\sigma\frac{\partial Q(p,s|x_{0})}{\partial x_{0}}\nonumber\\
 &+&\frac{1}{2}\sigma^{2}\frac{\partial^{2}Q(p,s|x_{0})}{\partial x_{0}^{2}}+...   ,
\end{eqnarray}
which upon inserting into Eq. (\ref{eq:bectrw3}) yields
\begin{eqnarray}
sQ(p,s|x_{0})-1&=&\frac{\sigma^{2}}{2}
K(s+pU(x_{0}))
\frac{\partial^{2}Q(p,s|x_{0})}{\partial x_{0}^{2}}\nonumber\\
&-&pU(x_{0})Q(p,s|x_{0}),\label{eq:bectrw4}
\end{eqnarray}
where
\begin{align}
K(s+pU(x_{0}))=\frac{[s+pU(x_{0})]\varphi(s+pU(x_{0}))}{1-\varphi(s+pU(x_{0}))}.
\end{align}
Finally, by performing Laplace inversion of Eq.\eqref{eq:bectrw4} with respect to
$t$, we obtain the following the backward equation 
\begin{widetext}
\begin{eqnarray}
   \frac{\partial Q(p,t|x_{0})}{\partial t}=\frac{\sigma^{2}}{2}\int_{0}^{t}K(t-t')e^{-p(t-t')U(x_{0})}\frac{\partial^{2}Q(p,t'|x_{0})}{\partial x_{0}^{2}}dt'
   -pU(x_{0})Q(p,t|x_{0}),
\label{FK2} 
\end{eqnarray}
\end{widetext}
as announced in the main text. 

\section{Derivation of Eq. \eqref{QTA}}
\label{Appendix-B}
We provide here the solution to the Feynman-Kac equation for the occupation time in an interval. First we Laplace transform Eq. \eqref{FK} with respect to time and get the ordinary differential equation 
\begin{eqnarray}
    sQ(p,s|x_{0})-1&=&\frac{\sigma^{2}}{2}K(s+pU(x_{0}))\frac{d^{2}Q(p,s|x_{0})}{d x_{0}^{2}}\nonumber\\
    &-&pU(x_{0})Q(p,s|x_{0}),
    \label{FK1}
\end{eqnarray}
where $U(x_0)=\theta (-a<x_0<a)$ and have considered the initial condition $T_a(t=0|x_{0})=0$
so that $Q(p,t=0|x_{0})=1$. In the region $x_0\in [-a,a]$, Eq. \eqref{FK1} turns into
\begin{align}
   \frac{\sigma^{2}}{2}K(s+p)\frac{d^{2}Q(p,s|x_{0})}{d x_{0}^{2}}-(s+p)Q(p,s|x_{0})=-1 ,
   \label{FK22}
\end{align}
while for $x_0\notin [-a,a]$, one should have
\begin{eqnarray}
    \frac{\sigma^{2}}{2}K(s)\frac{d^{2}Q(p,s|x_{0})}{d x_{0}^{2}}-sQ(p,s|x_{0})=-1.
    \label{FK3}
\end{eqnarray}
Finally, solving Eqs. \eqref{FK22} and \eqref{FK3} in each region we find
\begin{widetext}
\begin{eqnarray}
    Q(p,s|x_{0})=\left\{ \begin{array}{ll}
\frac{1}{s}+C_{1}e^{x_{0}\sqrt{\frac{2s}{\sigma^{2}K(s)}}}, & x_{0}<-a\\
\frac{1}{s+p}+C_{2}e^{x_{0}\sqrt{\frac{2(s+p)}{\sigma^{2}K(s+p)}}}+C_{3}e^{-x_{0}\sqrt{\frac{2(s+p)}{\sigma^{2}K(s+p)}}}, & -a<x_{0}<a\\
\frac{1}{s}+C_{4}e^{-x_{0}\sqrt{\frac{2s}{\sigma^{2}K(s)}}}, & x_{0}>a
\end{array}\right.
\end{eqnarray}
\end{widetext}
where we have taken into account the boundary conditions (see the main text Sec \ref{OT-interval}) $Q(p,s\big| |x_{0}|\rightarrow \infty)=1/s$. Since $Q(p,s|x_{0})=Q(p,s|-x_{0})$ from symmetry, we have $C_2=C_3=A$ and $C_1=C_4=B$. Demanding  continuity of the function $Q(p,s|x_{0})$ and its first derivative at $x_0=a$ we get the system of equations
$$
2A\cosh\left(\lambda(s+p)\right)-Be^{-\lambda(s)}=\frac{p}{s(s+p)},
$$
$$
2A\frac{\lambda(s+p)}{\lambda(s)}\sinh(\lambda(s+p))+Be^{-\lambda(s)}=0,
$$
where
$$
\lambda(s)=a\sqrt{\frac{2s}{\sigma^{2}K(s)}}.
$$
Solving for $A$ and $B$, and finally setting $x_0=0$, we find 
\begin{align}
    Q(p,s|x_{0}=0)=\frac{1}{s+p}+2A,
\end{align}
%$Q(p,s|x_{0}=0)=\frac{1}{s+p}+2A$ 
where
$$
A=\frac{p}{2s(s+p)}\frac{1}{\cosh\left(\lambda(s+p)\right)+\frac{\lambda(s+p)}{\lambda(s)}\sinh\left(\lambda(s+p)\right)}.
$$
This concludes the derivation of Eq. \eqref{QTA}.

\section{Derivation of Eqs. \eqref{TA1} and \eqref{TA2}}
\label{Appendix-C}
In here, we derive the exact expressions for the mean and second moment of the occupation time in the interval. We start by
inserting Eq. \eqref{FI} into the memory kernel in Eq. \eqref{eq:mk}
and by approximating for small $s$, to find
\begin{eqnarray}
 K(s)=\left\{ \begin{array}{ll}
\frac{s^{1-\alpha}}{b_{\alpha}}+..., & 0<\alpha<1\\
\frac{1}{\left\langle \tau\right\rangle }+\frac{b_{\alpha}s^{\alpha-1}}{\left\langle \tau\right\rangle ^2}+..., & 1<\alpha<2
\end{array}\right.   
\label{Kap}
\end{eqnarray}
and from Eq. \eqref{lambda}
\begin{eqnarray}
 \lambda(s)\simeq\left\{ \begin{array}{ll}
a\sqrt{\frac{2b_{\alpha}s^{\alpha}}{\sigma^{2}}}, & 0<\alpha<1\\
a\sqrt{\frac{2\left\langle \tau\right\rangle s}{\sigma^{2}}}\left(1-\frac{b_{\alpha}}{2\left\langle \tau\right\rangle }s^{\alpha-1}\right), & 1<\alpha<2
\end{array}\right..
\label{lap}
\end{eqnarray}
Substituting this expression into Eq. \eqref{m1l} gives us
\[
\left\langle T_{a}(s)\right\rangle \simeq\left\{ \begin{array}{ll}
a\sqrt{\frac{2b_{\alpha}}{\sigma^{2}}}\frac{1}{s^{2-\alpha/2}}, & 0<\alpha<1\\
a\sqrt{\frac{2\left\langle \tau\right\rangle }{\sigma^{2}}}\left(\frac{1}{s^{3/2}}+\frac{b_{\alpha}}{2\left\langle \tau\right\rangle s^{5/2-\alpha}}\right), & 1<\alpha<2
\end{array}\right.
\]
 where we have approximated for small $s$ again. Applying the Laplace
inversion to this result we readily find Eq. \eqref{TA1}.

For the second moment, we plug Eq. \eqref{Kap} into Eq. \eqref{Fi} to have
\begin{eqnarray}
    \Phi(s)\simeq\left\{ \begin{array}{ll}
\frac{\alpha}{2}, & 0<\alpha<1\\
\frac{1}{2}\left[1+(1-\alpha)\frac{b_{\alpha}}{\left\langle \tau\right\rangle }s^{\alpha-1}\right], & 1<\alpha<2
\end{array}\right.
\end{eqnarray}
which upon insertion to Eq. \eqref{m2l} provides
\begin{eqnarray*}
\left\langle T_{a}(s)^{2}\right\rangle \simeq\left\{ \begin{array}{ll}
2a\sqrt{\frac{2b_{\alpha}}{\sigma^{2}}}\frac{1-\alpha}{s^{3-\alpha/2}}+\frac{a^{2}(3\alpha-1)}{s^{3-\alpha}}\frac{2b_{\alpha}}{\sigma^{2}}, & 0<\alpha<1\\
\frac{2a^{2}}{Ds^{2}}+\frac{2a(\alpha-1)b_{\alpha}}{\left\langle \tau\right\rangle \sqrt{D}}\frac{1}{s^{7/2-\alpha}}, & 1<\alpha<2
\end{array}\right.
\end{eqnarray*}
where $D=\sigma ^2/2\left\langle \tau\right\rangle$. Performing the inverse Laplace transform to the above result we obtain
\begin{widetext}
 \begin{eqnarray}
\left\langle T_{a}(t)^2\right\rangle \simeq\left\{ \begin{array}{ll}
2a\sqrt{\frac{2b_{\alpha}}{\sigma^{2}}}\frac{1-\alpha}{\Gamma(3-\alpha/2)}t^{2-\alpha/2}+\frac{a^{2}(3\alpha-1)}{\Gamma(3-\alpha)}\frac{2b_{\alpha}}{\sigma^{2}}t^{2-\alpha}, & 0<\alpha<1\\
\frac{2a^{2}}{D}t+\frac{2a(\alpha-1)b_{\alpha}}{\Gamma\left(7/2-\alpha\right)\left\langle \tau\right\rangle \sqrt{D}}t^{\frac{5}{2}-\alpha}, & 1<\alpha<2.
\end{array}\right. .
\end{eqnarray}   
\end{widetext}

For $0<\alpha<1$ the second term of the right hand side is much lower than the first term in the large time limit ($s\to 0$) so that it can be removed. This leads to Eq. \eqref{TA2}.

\section{Derivation of Eqs. \eqref{QTABM} and \eqref{QTAT0}}
\label{Appendix-D}
In this appendix, we provide the details of the derivation for the PDF of occupation time in the interval. We start by deriving Eq. \eqref{QTABM} first. 

Recalling expressions from Eqs \eqref{fiapfm} and \eqref{lambda}, we have $\lambda (s)\simeq a\sqrt{2s\left\langle \tau\right\rangle/\sigma^2}$ as $s\to 0$, so that Eq. \eqref{qps1} turns into
\begin{align}
    Q(p,s)\simeq\frac{1}{p}+\frac{2}{\sqrt{sp}}e^{-\frac{a}{\sqrt{D}}\sqrt{p}}.
\end{align}
To invert this expression in Laplace with respect to $s$ we use
$$
\mathcal{L}_{s\to t}^{-1}\left[\frac{1}{\sqrt{s}}\right]=\frac{1}{\sqrt{\pi t}},
$$
so that
$$
Q(p,t)\simeq\frac{\delta(t)}{p}+\frac{2}{\sqrt{\pi pt}}e^{-\frac{a}{\sqrt{D}}\sqrt{p}}.
$$
Next we invert the above expression with respect to $p$ using
$$
\mathcal{L}_{p\to T_{a}}^{-1}\left[\frac{e^{-\frac{a}{\sqrt{D}}\sqrt{p}}}{\sqrt{p}}\right]=\frac{e^{-\frac{a^{2}}{4DT_{a}}}}{\sqrt{\pi T_{a}}},
$$
and find
\begin{eqnarray*}
    Q(T_{a},t)\simeq \frac{2}{\pi\sqrt{tT_{a}}}e^{-\frac{a^{2}}{4DT_{a}}}.
\end{eqnarray*}
which is Eq. \eqref{QTABM}.
\\

To derive Eq. \eqref{QTAT0} we make use of Eq. \eqref{lap} derived in the previous appendix. This recalls $\lambda(s)=as^{\alpha/2}/\sqrt{K_{\alpha}}$ where $K_\alpha = \sigma^2/2b_\alpha$, plugging the same into Eq. \eqref{qps1} we find
$$
Q(p,s)\simeq\frac{1}{p}+\frac{2e^{-\frac{ap^{\alpha/2}}{\sqrt{K_{\alpha}}}}}{s^{1-\frac{\alpha}{2}}p^{\alpha/2}}.
$$
Inverting this expression with respect to $s$ using
$$
\mathcal{L}_{s\to t}^{-1}\left[\frac{1}{s^{1-\frac{\alpha}{2}}}\right]=\frac{t^{-\alpha/2}}{\Gamma\left(1-\frac{\alpha}{2}\right)},
$$
we find
\begin{eqnarray}
    Q(p,t)\simeq\frac{\delta(t)}{p}+\frac{2e^{-\frac{ap^{\alpha/2}}{\sqrt{K_{\alpha}}}}}{\Gamma\left(1-\frac{\alpha}{2}\right)(pt)^{\alpha/2}}.
    \label{aux}
\end{eqnarray}
Now, we invert with respect to $p$ by noticing that the one-sided L\'evy density has the characteristic function given by Eq. \eqref{os}, and so
$$
\mathcal{L}_{p\to T_{a}}^{-1}\left[e^{-\frac{ap^{\alpha/2}}{\sqrt{K_{\alpha}}}}\right]=\left(\frac{\sqrt{K_{\alpha}}}{a}\right)^{\frac{2}{\alpha}}l_{\alpha/2}\left[\left(\frac{\sqrt{K_{\alpha}}}{a}\right)^{\frac{2}{\alpha}}T_{a}\right].
$$
Hence,
$$
\mathcal{L}_{p\to T_{a}}^{-1}\left[\frac{e^{-\frac{ap^{\alpha/2}}{\sqrt{K_{\alpha}}}}}{p^{\alpha/2}}\right]=\frac{\left(\sqrt{K_{\alpha}}/a\right)^{\frac{2}{\alpha}}}{\Gamma\left(\alpha/2\right)}\int_{0}^{T_{a}}\frac{l_{\alpha/2}\left[\left(\frac{\sqrt{K_{\alpha}}}{a}\right)^{\frac{2}{\alpha}}t'\right]}{\left(T_{a}-t'\right)^{1-\frac{\alpha}{2}}}dt'.
$$
Using the above result, the inverse Laplace transform of Eq. \eqref{aux} results in Eq. \eqref{QTAT0}. A series expansion can be done in the following way that might be useful to evaluate the expression
\begin{align}
Q(T_a,t)\simeq\frac{2}{\Gamma\left(1-\frac{\alpha}{2}\right)t^{\frac{\alpha}{2}}T_{a}^{1-\frac{\alpha}{2}}}\sum_{n=0}^{\infty}\frac{\left(-\frac{aT_{a}^{-\frac{\alpha}{2}}}{\sqrt{K_{\alpha}}}\right)^{n}}{n!\Gamma\left(\frac{\alpha}{2}-\frac{\alpha n}{2}\right)}. 
\end{align}

%expressed in terms of a Fox function. Using (see Eq. 2.23 in Ref. \cite{Mat09}) in \eqref{Q35}
%\begin{eqnarray}
%    \mathcal{L}_{p\to z}^{-1}\left[\frac{e^{-Ap^{\gamma}}}{p^{\rho}}\right]=z^{\rho-1}H_{1,1}^{1,0}\left[Az^{-\gamma}\left|\begin{array}{c}
%(\rho,\gamma)\\
%(0,1)
%\end{array}\right.\right]
%\label{fox1}
%\end{eqnarray}

\section{Derivation of Eq. \eqref{QTAT022}}
\label{Appendix-E}
To get the asymptotic expression for the PDF of occupation time in the right edge regime, we
insert $K(s)$ given in Eq. \eqref{Kap} for $0<\alpha<1$ into Eq. \eqref{Qpe}. This yields
\begin{eqnarray}
 Q_{\epsilon}(p_{\epsilon},s)\sim\frac{1}{s}-\frac{1}{s}\frac{p_{\epsilon}^{\alpha/2}}{p_{\epsilon}^{\alpha/2}+\frac{as^{\alpha}}{\sqrt{K_{\alpha}}}},   
\end{eqnarray}
which yields
\begin{eqnarray}
    Q_{\epsilon}(p_{\epsilon},s)\sim\frac{\delta(\epsilon)}{s}-\frac{1}{s\epsilon}E_{\frac{\alpha}{2},0}\left(-\frac{as^{\alpha}\epsilon^{\alpha/2}}{\sqrt{K_{\alpha}}}\right),
\end{eqnarray}
after the Laplace inversion with respect to $p_\epsilon$. Here, $E_{a,b}(z)$ is two-parametric Mittag-Leffler function \cite{Go20}. Since we are interested in the small $s$ and $\epsilon$ limits, we use the small argument expansion of the Mittag-Leffler function (see Eq. 4.1.1 in Ref. \cite{Go20}). This leads to
\begin{eqnarray}
    Q_{\epsilon}(\epsilon,s)\sim\frac{\delta(\epsilon)}{s}+\frac{as^{\alpha-1}\epsilon^{\frac{\alpha}{2}-1}}{\sqrt{K_{\alpha}}\Gamma(\alpha/2)}.
    \label{QTAT02}
\end{eqnarray}
The Laplace inversion of the above expression to real time leads to Eq. \eqref{QTAT022} as in the main text.

\section{A brief primer on \textit{Ergodicity}}
\label{Appendix-F}
Let us consider a stochastic trajectory $x(\tau)$  observed from $\tau=0$ up to time $\tau = t$. Consider an observable $\mathcal{O}[x(\tau)]$, a function of the trajectory $x(\tau)$. Since $x(\tau)$ is stochastic in nature, the observable $\mathcal{O}[x(\tau)]$ will also be fluctuating between the realizations. An observable of the random walk is said to be ergodic if the ensemble average equals the time average $\left\langle \mathcal{O}\right\rangle =\ensuremath{\overline{\mathcal{O}}}$ in the long time limit. This means that if $\mathcal{O}[x(\tau)]$ is ergodic then its time average $\ensuremath{\overline{\mathcal{O}}}$ is not a random variable. As a consequence, the limiting PDF of $\ensuremath{\overline{\mathcal{O}}}$ is \begin{eqnarray}
Q(\overline{\mathcal{O}},t\to\infty)=\delta\left(\overline{\mathcal{O}}-\left\langle \mathcal{O}\right\rangle \right).
     \label{lpdf}
 \end{eqnarray} 
 At this point, let us define the density $P(x,t)$ which is the probability to find the walker at the point $x$ at time $t$, i.e., it is the propagator.
 If the observable is integrable with respect to the density $P(x,t)$, then the ensemble average is
\begin{equation}
    \left\langle \mathcal{O}[x(t)]\right\rangle =\int_{-\infty}^{\infty}\mathcal{O}[x]P(x,t)dx.
\end{equation}
The time average of $\mathcal{O}[x(t)]$ is defined as
\begin{equation}
    \ensuremath{\overline{\mathcal{O}[x(t)]}=}\frac{1}{t}\int_{0}^{t}\mathcal{O}[x(\tau)]d\tau.
\end{equation}
For non-ergodic observables, since $\overline{\mathcal{O}}$ is random, its variance $\textrm{Var}(\overline{\mathcal{O}})$ is non-zero in the long time limit. Otherwise, for an ergodic observable $\textrm{Var}(\overline{\mathcal{O}})=0$ in the long time limit. Keeping this in mind, one can define the ergodicity breaking parameter EB
in the following way
\begin{eqnarray}
\textrm{EB}=\lim_{t\to\infty}\frac{\textrm{Var}(\overline{\mathcal{O}})}{\left\langle \overline{\mathcal{O}}\right\rangle ^{2}}=\lim_{t\to\infty}\frac{\left\langle \overline{\mathcal{O}}^{2}\right\rangle -\left\langle \overline{\mathcal{O}}\right\rangle ^{2}}{\left\langle \overline{\mathcal{O}}\right\rangle ^{2}}.
  \label{EB}
\end{eqnarray}
For ergodic observables, one should have $\textrm{EB}= 0$.

\section{Derivation of Eq. \eqref{QTP}}
\label{Appendix-G}
In this appendix, we outline the derivation of the MGF for the half-occupation time. To do this, 
we need solve Eq. \eqref{FK1} with the function $U(x_0)=\theta (x_0)$ in the regions $x_0>0$ and $x_0<0$ requiring continuity at $x_0=0$. Thus, 
Eq. \eqref{FK1} splits into

\[
\frac{\sigma^{2}}{2}K(s+p)\frac{d^{2}Q(p,s|x_{0})}{dx_{0}^{2}}-(s+p)Q(p,s|x_{0})=-1,
\]
for $x_{0}>0$ and

\[
\frac{\sigma^{2}}{2}K(s)\frac{d^{2}Q(p,s|x_{0})}{dx_{0}^{2}}-sQ(p,s|x_{0})=-1,
\]
for $x_{0}<0$. Applying the boundary conditions $\ensuremath{Q(p,s|x_{0}\rightarrow+\infty)=1/(s+p)}$
and $Q(p,s|x_{0}\rightarrow-\infty)=1/s$ (see Sec \ref{OT-half-interval}), the general solution can 
be written as
\begin{align}
  Q(p,s|x_{0}>0)&=\frac{1}{s+p}+C_{1}e^{-x_{0}\sqrt{\frac{2(s+p)}{\sigma^{2}K(s+p)}}} , \\
  Q(p,s|x_{0}<0)&=\frac{1}{s}+C_{2}e^{x_{0}\sqrt{\frac{2s}{\sigma^{2}K(s)}}},
  \label{Q-HI-sol}
\end{align}
where the integration constants can be determined by imposing continuity of the function
$Q(p,s|x_{0})$ and its derivative with respect to $x_{0}$ at $x_{0}=0.$ This leads to the following set of equations
$C_{1}$ and $C_{2}$
\begin{eqnarray}
  C_{1}-C_{2}&=&\frac{p}{s(s+p)},\\
  C_{2}&=&-C_{1}\sqrt{\frac{(s+p)K(s)}{sK(s+p)}}.
\end{eqnarray}
that can be solved easily. Plugging back into Eq. \eqref{Q-HI-sol}, the final solution at $x_{0}=0$ can be found as announced in Eq. \eqref{QTP} of the main text. 

\section{Derivation of the limiting distributions Eq. \eqref{Qzl} and Eq. \eqref{lamperti}}
\label{Appendix-H}
In Sec \ref{OT-half-interval}, we have highlighted the limiting forms of the scaled half-occupation time PDFs namely Eq. \eqref{Qzl} and Eq. \eqref{lamperti}. Here, we outline the steps leading to these limiting forms. 
If the moment generating function $Q(p,s)$ has the following scaling behavior 
$$
Q(p,s)=\frac{1}{s}g\left(\frac{p}{s}\right),
$$
in the limits $s\rightarrow 0$ and $p\rightarrow 0$ with $p/s$ arbitrary then the random variable $y=T^+/t$ possesses a limiting distribution 
$$
Q(y)=\lim_{t\rightarrow\infty}Q\left(y=\frac{T^+}{t},t\right)=\lim_{t\rightarrow\infty}tQ\left(T^+=yt,t\right)
$$
which can be computed as was done in \cite{GoLu01} by noting
\begin{eqnarray}
Q(y)=-\frac{1}{\pi y}\lim_{\epsilon\rightarrow0}\textrm{Im}\left[ g\left(-\frac{1}{y+i\epsilon}\right)\right].
\label{ld}
\end{eqnarray}
From Eq. \eqref{eq:mk} we see that the limit of $s/K(s)$ is
$$
\lim_{s\rightarrow0}\frac{s}{K(s)}=\lim_{s\rightarrow0}\left[\frac{1}{\varphi(s)}-1\right]=0,
$$
for any $\varphi(s)$ or $K(s)$, i.e, for any random walk. Thus, from Eq. \eqref{QTP} the scaling function is given by
\begin{eqnarray}
  g(\chi)=\frac{1}{1+\chi}\left[1+\frac{\chi}{1+\sqrt{(1+\chi)f(\chi)}}\right] , 
  \label{gchi}
\end{eqnarray}
where $\chi=p/s$ and $f(\chi)=\lim_{\{s,p\}\rightarrow0}\frac{K(s)}{K(s+p)}$. If the waiting time PDF has finite moments, from Eq. \eqref{fiapfm} one finds $K(s)=\left\langle t\right\rangle ^{-1}+O(s)$ so that $f(\chi)=1$. Inserting this into Eq. \eqref{gchi} and computing the limit in Eq. \eqref{ld} we get
$$
Q(y)=\frac{1}{\pi\sqrt{y(1-y)}},\quad 0<y<1.
$$
The above expression is the \textit{arcsine distribution} as was mentioned in the main text. 
If we consider the ML waiting time PDF \eqref{FII} then one finds $f(\chi)=(1+\chi)^{-1+\alpha}$ and the limiting distribution is given by the Lamperti distribution \cite{La58,CaBa10}
$$
Q(y)=\frac{\sin(\frac{\alpha\pi}{2})}{\pi}\frac{y^{\frac{\alpha}{2}-1}(1-y)}{y^{\alpha}+(1-y)^{\alpha}+2y^{\frac{\alpha}{2}}(1-y)\cos(\frac{\alpha\pi}{2})},
$$
which is Eq. \eqref{lamperti} from the main text.

\section{Derivation of $\left\langle T^{+}(t)^{n}\right\rangle$ }
\label{Appendix-I}
This appendix provides detailed derivation for the $n$-th order moment of the half occupation time $\left\langle T^{+}(t)^{n}\right\rangle$ which was discussed in Sec \ref{OT-half-interval}A. 

To this end, we recall the expression for $Q(p,s)$ from Eq. \eqref{QTP} which can be written in the form
\[
Q(p,s)=\frac{1}{s}\frac{u+u^{\alpha/2}}{1+u^{\alpha/2}},
\]
with $u=s/(s+p).$ Since $u<1$ we can expand the denominator as the power
series
\[
\frac{1}{1+u^{\alpha/2}}=\sum_{j=0}^{\infty}(-1)^{j}u^{\alpha j/2},
\]
so that $Q(p,s)$ can be expressed as
\begin{equation}
Q(p,s)=\frac{1}{s}\sum_{j=0}^{\infty}(-1)^{j}\left[u^{1+\alpha j/2}+u^{\alpha(1+j)/2}\right].\label{qtn}
\end{equation}
To compute $\left\langle T^{+}(s)^{n}\right\rangle $ we make use
of the relation \eqref{moments2} and expression in Eq. \eqref{qtn} leading to
\begin{eqnarray}
\left(\frac{\partial^{n}Q(p,s)}{\partial p^{n}}\right)_{p=0}&=&\frac{1}{s}\sum_{j=0}^{\infty}(-1)^{j}\left[\frac{\partial^{n}}{\partial p^{n}}\left(u^{1+\alpha j/2}\right)_{p=0}\right.\nonumber\\
&+&\left.\frac{\partial^{n}}{\partial p^{n}}\left(u^{\alpha(1+j)/2}\right)_{p=0}\right].\label{eq:Qpn}
\end{eqnarray}
To find the expression for the $n-$th derivatives with respect to
$p$, we can write the derivative of the power $u^{\mu}$ (with $\mu>0$) as
\begin{align}
\frac{\partial^{n}}{\partial p^{n}}\left[u^{\mu}\right]_{p=0}=\frac{\partial^{n}}{\partial p^{n}}\left[\left(\frac{s}{s+p}\right)^{\mu}\right]_{p=0}=\frac{(-1)^{n}}{s^{n}}\frac{\Gamma\left(n+\mu\right)}{\Gamma\left(\mu\right)}. \end{align}
Making use of the above formula, Eq. \ref{eq:Qpn} turns to
\begin{eqnarray}
  \left(\frac{\partial^{n}Q(p,s)}{\partial p^{n}}\right)_{p=0}&=&\frac{(-1)^{n}}{s^{1+n}}\sum_{j=0}^{\infty}(-1)^{j}\left[\frac{\Gamma\left(n+1+\frac{\alpha j}{2}\right)}{\Gamma\left(1+\frac{\alpha j}{2}\right)}\right.\nonumber\\
  &+&\left.\frac{\Gamma\left(n+\frac{\alpha}{2}+\frac{\alpha j}{2}\right)}{\Gamma\left(\frac{\alpha}{2}+\frac{\alpha j}{2}\right)}\right]  .
\end{eqnarray}
The above expression can be inserted into Eq. \eqref{moments2} to arrive at our desired expression in Eq. \eqref{tnht}.

\bibliography{main}
\end{document}